\documentclass{aa}
\usepackage{graphicx}
\usepackage{natbib}
\bibpunct{(}{)}{;}{a}{}{,}

\begin{document}

\newcommand{\reduceme}{\mbox{R\raisebox{-0.35ex}{E}D%
\hspace{-0.05em}\raisebox{0.85ex}{uc}\hspace{-0.90em}%
\raisebox{-.35ex}{{m}}\hspace{0.05em}E}}
\newcommand{\onen}{\makebox[0in][r]{1}}

\title{Minor axis kinematics of 19 S0-Sbc bulges}

\author{J. Falc\'on-Barroso\inst{1} 
\and M. Balcells\inst{2} 
\and R.F. Peletier\inst{1,3}
\and A. Vazdekis\inst{2}}

\institute{School of Physics \& Astronomy. University of Nottingham. 
Nottingham. NG7 2RD. United Kingdom. \\
\email{ppxjf2@nottingham.ac.uk}\\
\email{Reynier.Peletier@nottingham.ac.uk}\and
Instituto de Astrof\'\i sica de Canarias, E-38200 La Laguna, Tenerife, Spain. \\
\email{balcells@ll.iac.es}\\
\email{vazdekis@ll.iac.es}\and
CRAL, Observatoire de Lyon, F-69561 St-Genis Laval cedex, France. \\}

\date{Received ...; accepted ...}

\abstract{We present minor axis kinematic profiles for a well-studied sample of
19 early- to intermediate-type disk galaxies.  We introduce, for the first time, 
the use of single-burst stellar population (SSP) models to obtain stellar velocities,
velocity dispersions and higher order Gauss-Hermite moments ($h_3$, $h_4$) from
galaxy spectra in the near-infrared Ca\,{\sc ii} triplet region.  SSP models,
which employs the synthetic spectra of \citet{vazdekis03}, provide a means to address
the template-mismatch problem, and are shown to provide as good or better fits
as traditional stellar templates.   We anticipate the technique to be of
particular use for high-redshift galaxy kinematics.  We give the measurement of
a recently defined CaT$^*$ index \citep{cenarro01}, and describe the global
properties of the bulge kinematics as derived from the kinematic profiles.  We
detect small-amplitude minor-axis rotation, generally due to inner isophotal
twists as a result of slightly triaxial bulges or misaligned inner disks; such
inner features do not show peculiar colors or distinct CaT$^*$ index values.
Velocity dispersion profiles, which extend well into the disk region, show a
wide range of slopes.  Flattened bulges tend to have shallower velocity
dispersion profiles.  The inferred similarity of bulge and disk {\it radial}
velocity dispersions supports the interpretation of these bulges as thickened
disks.   
}

\maketitle

\section{Introduction}
\label{Sec:Intro}

The measurement of kinematics in galaxies has been one of the major achievements
in astrophysics in the second half of the last century.  Our knowledge in this
field has evolved from crude kinematic determinations from photographic plates 
\citep{b68}, to high-resolution, detailed measurements of the line-of-sight 
velocity distributions (hereafter LOSVDs; i.e. \citealt{bender94,f97,JVB01}). 

In the seventies, methods were devised to measure velocity dispersions 
in addition to radial velocities, i.e. cross-correlation \citep{tonry79},
Fourier Quotient \citep{sargent77}. In the  past decade, with the improved
quality of the data, a new set of  algorithms were developed allowing us to
measure LOSVDs (Fourier  Correlation Quotient, \citealt{bender90}), and
quantify deviations from pure single Gaussian \citep{vdm93} and multiple
Gaussian LOSVD with UGD \citep{km93,prada96}.  Galactic kinematics, however, 
suffers from an important technical constraint: template mismatching. Stars 
are observed together with the object which kinematics we are interested in, to 
be used as spectral templates.  However, it is difficult to choose the right star, 
since galaxy light comes from a  wide range of stellar types.  Although, for 
ellipticals in the  $V$-band, it has been found that half of the light comes from 
G and  K giants \citep{pickles85} and, therefore, those stellar types are good 
templates, this situation may change dramatically from galaxy to galaxy  or even 
within a galaxy. Therefore kinematic parameters will be  sensitive to the stellar 
template used. Much work has been devoted to address the template mismatch problem
\citep{rw92,g93,ssc96,ss99}. 

In the last decade, new evolutionary stellar population synthesis models
\citep{bruzual93,worthey94,buzz95,tan96,vazdekis96} have helped us to
significantly increase our galaxy stellar populations knowledge. These models
predict line strengths of a limited number of spectral features, such as those
of the Lick/IDS system (i.e.\citealt{worthey94,vazdekis96}). However since
these models were unable to predict full SEDs (Spectral Energy Distributions) at
high enough resolution, they could not be used as templates for galaxy stellar
kinematics. This situation has recently changed with the development
of new SSP model SEDs in the blue and visual at resolution 1.8~\AA~(FWHM) by
\citet{vazdekis99} and in the near-IR Ca\,{\sc ii} triplet region at 1.5~\AA\
\citep{vazdekis03}. In fact, \citet{vazdekis99} and \citet{vazari99} have shown the
potential use of the SSP models as templates for determining the main kinematic
parameters in the optical spectral range. In this paper we are for the first
time in position to determine galaxy kinematics using SSP templates in the
near-IR Ca\,{\sc ii} spectral range. It is generally assumed, that kinematic
stellar templates are required to be obtained with the same instrumental 
setting as the one used for the galaxy spectra \citep{sargent77,tonry79}. 
In this paper we show that this is no longer a requirement, at
least at the resolutions of the spectra considered here.

This paper intends to be a first step in the exploration of SSP models as
templates by comparing velocity profiles, for a sample of 19 galaxies, with
those obtained with our best stellar or mixture of stellar templates. We show 
that in all cases SSP models produce kinematics of at least at the same level 
of quality, and are quite often better than the best stellar templates. In order 
to trace correlations between kinematic and  population features along the galaxy minor
axes, we measure the CaT$^*$ index \citep{cenarro01}.

Sect.~2 summarizes the observations and data reduction. In Sect.~3 we describe
the different types of template samples and the way the kinematical profiles
have been determined, including a detailed comparison between stellar templates
and SSP models. The obtained kinematics is discussed in Sect.~4. Kinematical
profiles are graphically presented in Appendix~\ref{Ap:ProfilePlots}. Tables 
containing the numerical values presented in appendix~\ref{Ap:ProfilePlots} will 
be available in electronic form.

\section{Observations \& Data Reduction}

\subsection{Sample}
\label{Sub:Sample}

The sample comprises 20 galaxies of types S0--Sbc selected out of the 
\citet{bp94} sample of bulges.  The latter is a diameter-limited sample 
of galaxies with inclinations above 50 degree from the UGC catalog 
\citep{nilson95}; it is has been the subject of several studies on bulges 
(see \citealt{fpb02} and references therein).  The 20 objects were those 
we had observed with HST/WFPC2 and NICMOS, hence high-spatial resolution 
images are available for them at $B,I,H$ bands \citep{pb99}. The main 
properties of the sample are presented in Table~\ref{Tab:Sample}.

\subsection{Observations}
\label{Sub:Observations}

We obtained long-slit spectra for our sample of galaxies at the 4.2m William
Herschel Telescope of the Observatorio del Roque de los Muchachos at La Palma
between 11 and 13 July 1997. Twenty galaxies  were observed between 8360 -
9170~\AA\ using the ISIS double spectrograph.  We discarded NGC~5577 as the
spectra did not have sufficient  signal-to-noise. We are thus left with 19
objects (see Table~\ref{Tab:Sample}).  The red arm was equipped with the
Tektronix (1024$\times$1024) TEK2 CCD (0.36\arcsec per 24 $\mu$m pixel) and the
R600R grating, providing a spectral resolution (FWHM of arc lines) of
1.74~\AA\ (59.6 km/s) with a spectral dispersion of 0.8 \AA/pix. We used the
6100~\AA\ dichroic for simultaneous red and blue arm exposures. In this paper
we discuss the red arm spectra only.  Typical exposure times per galaxy were 
1500~sec. The length of the slit was 3.6\arcmin. The slit width was 
1.2\arcsec, matching the seeing at the time of the observations. Arc line exposures 
were taken before and after each target exposure. Tungsten continuum lamp 
exposures were taken with the red arm after each target exposure, for fringe 
calibration. Twilight sky exposures were taken every night for flat fielding. 
We also obtained spectra for 11 stellar templates of types A2 to M4, from the 
Lick list of stars \citep{worthey94}, listed in Table~\ref{Tab:templates}.

\begin{table*}
\begin{center}
\caption{\sc Galaxy Sample}
\label{Tab:Sample}
{\tabcolsep=4.3pt
\begin{tabular}{cccrcclrcllccc}
\hline\hline
  NGC   & V$_{\rm{LG}}$  &    Scale    & Type &  M$_R^{tot}$  &  M$_R^{bul}$  &  $B/D$   &  PA     & Exp. Time & \Large{$\epsilon$}\tiny{$_D$}  & \Large{$\epsilon$}\tiny{$_B$} &     r$_{bulge}$  & Best Templ.& S/N\\
   ~    &      km/s      & kpc/\arcsec &   ~  &  mag	   &	  mag	      &      ~   &   deg   &   sec     &      ~ 		        &	     ~  		&        \arcsec   & ~   	& pix$^{-1}$\\
  (1)   &      (2)       &     (3)     &  (4) &  (5)	   &	  (6)	      &    (7)   &   (8)  &   (9)    &    (10) 		        &	   (11)  		&      (12)        & (13)	&  (14)\\
\hline
  5326  & 2576  	 & 0.25        &  1   &  $-$22.16  & $-$21.22	      &  0.73	 & -138    & 1500      &  0.50  		        &  0.43  			&  5.97            & $+$0.2, 3.16    	     &  83\\
  5389  & 1990  	 & 0.19        &  0   &  $-$21.32  & $-$20.70	      &  1.30	 &   93    & 1500      &  0.80  		        &  0.46  			&  2.85            & M1 III / $+$0.2, 4.47   &  61\\
  5422  & 1929  	 & 0.19        & $-$2 &  $-$21.96  & $-$21.33	      &  1.28	 &   64    & 1500      &  0.80  		        &  0.35  			&  3.33            & $+$0.2, 7.08    	     &  68\\
  5443  & 2060  	 & 0.20        &  3   &  $-$21.53  & $-$19.09	      &  0.12	 &  -52    & 1400      &  0.68  		        &  0.54  			&  1.98            & $+$0.2, 3.16    	     &  37\\
  5475  & 1861  	 & 0.18        &  0   &  $-$20.98  & $-$18.72	      &  0.14	 & -101    & 1500      &  0.68  		        &  0.24  			&  2.45            & $+$0.2, 1.41    	     &  57\\
  5587  & 2291  	 & 0.22        &  0   &  $-$21.07  & $-$19.43	      &  0.29	 &   73    & 1500      &  0.70  		        &  0.47  			&  1.51            & $-$0.4, 14.12   	     &  36\\
  5689  & 2290  	 & 0.22        &  0   &  $-$22.29  & $-$21.47	      &  0.89	 &   -5    & 1500      &  0.75  		        &  0.48  			&  2.04            & 0.0, 4.47       	     &  58\\
  5707  & 2354  	 & 0.23        &  2   &  $-$21.43  & $-$20.11	      &  0.42	 &  -59    & 1500      &  0.75  		        &  0.20  			&  3.79            & 0.0, 2.51      	     &  51\\
  5719  & 1684  	 & 0.16        &  2   &  $-$21.60  & $-$20.00	      &  0.30	 &   5     & 1500      &  0.64  		        &  0.61  			&  7.71            & K3 Ib / $-$0.4, 3.55    &  51\\
  5746  & 1677  	 & 0.16        &  3   &  $-$22.68  & $-$21.70	      &  0.68	 &  -99    & 1200      &  0.84  		        &  0.52  			&  2.37            & M1 III / $+$0.2, 3.16   &  44\\
  5838  & 1337  	 & 0.13        & $-$3 &  $-$21.89  & $-$20.93	      &  0.71	 &  134    & 1500      &  0.65  		        &  0.22  			&  8.39            & 0.0, 12.59      	     &  86\\
  5854  & 1708  	 & 0.17        & $-$1 &  $-$21.43  & $-$20.18	      &  0.46	 &  -33    & 1500      &  0.70  		        &  0.32  			&  4.85            & $-$0.7, 6.31    	     &  64\\
  5879  & 1065  	 & 0.10        &  4   &  $-$20.30  & $-$19.08	      &  0.48	 &   89    & 1500      &  0.62  		        &  0.32  			&  0.85            & $-$0.4, 3.16    	     &  41\\
  5965  & 3603  	 & 0.35        &  3   &  $-$22.92  & $-$21.76	      &  0.53	 &  -38    & 1500      &  0.84  		        &  0.50  			&  2.32            & $-$0.4, 5.01    	     &  46\\
  6010  & 1923  	 & 0.19        &  0   &  $-$21.57  & $-$19.88	      &  0.27	 &   12    & 1500      &  0.75  		        &  0.28  			&  2.70            & 0.0, 14.12      	     &  67\\
  6504  & 4680  	 & 0.46        &  2   &  $-$24.57  & $-$23.30	      &  0.45	 &  135    & 1500      &  0.83  		        &  0.44  			&  4.05            & $-$0.4, 3.16    	     &  52\\
  7331  & 1138  	 & 0.11        &  3   &  ---	   & ---	      & 0.001$^1$&   80    & 1500      &  0.63$^1$		        &  0.62  			&  4.57            & $-$0.4, 15.85   	     & 117\\
  7332  & 1550  	 & 0.15        & $-$2 &  $-$21.86  & $-$20.52	      &  0.41	 & -114    & 1500      &  0.74  		        &  0.29  			&  2.99            & $-$0.4, 11.22   	     & 101\\
  7457  & 1114  	 & 0.11        & $-$3 &  $-$20.91  & $-$20.91	      & 10.00	 &  35     & 1500      &  0.48  		        &  0.34  			&  3.29            & $+$0.2, 2.51    	     &  51\\
\hline\hline
\end{tabular}}
\end{center}
{\bf Notes:} Description of the columns:
\newline (1): the NGC numbers of the galaxies.
\newline (2): Recession velocity of each galaxy in km/s, corrected to the Local Group 
\citep{kk96}, from optical heliocentric velocities given in the {\it Third Reference 
Catalogue of Bright Galaxies} \citeyearpar{RC3} (hereafter RC3).
\newline (3): Scale in kpc/arcsec from \citet{fpb02}.
\newline (4): Type index $T$ from the RC3.
\newline (5) and (6): Absolute magnitudes, in $R$-band, from \citet{pb97} (H$_0$=50 km/s/Mpc).
\newline (7): Bulge-to-disk ratio from an $R$-band bulge-disk decomposition 
	following Kent's \citeyearpar{Kent84} method \citep{pb97}.
\newline (8): The position angle, in degrees, (N -- E) of the dust-free minor axis \citep{pb97}.
\newline (9): Integration time, in sec, for each galaxy during the observations.
\newline (10): Disk ellipticity from \citet{pb97}.
\newline (11): Effective bulge ellipticity, from HST $H$-band ellipticity profiles, measured 
at the geometric effective radius \citep{balcells03}.
\newline (12): Minor axis bulge radius, derived from \citet{balcells03} HST $H$-band profiles,
as the radius at which the contribution from the disk and bulge light are equal, except for
NGC~7332 for which a ground-based image was used instead \citep{pb97} (Alister W. Graham, private
communication).
\newline (13) Best template fit for each galaxy. The metallicity ([m/H]) and age
(Gyr) of the best fit SSP models is shown. In those cases where the stellar 
template performed better, the adopted SSP best template is also given. 
\newline (14) Signal-to-noise ratio per pixel in the central aperture (1.2 x 0.36 arcsec$^2$).
\newline $^1$ From \citet{prada96}
\end{table*}

\subsection{Data Reduction}
\label{Sub:Data Reduction}
The data reduction was carried out using the IRAF\footnote{IRAF is distributed by 
    the National Optical Astronomy Observatories,
    which are operated by the Association of Universities for Research
    in Astronomy, Inc., under cooperative agreement with the National
    Science Foundation.}
package. The general standard procedures were applied to the data (bias 
subtraction, flat fielding, fringe correction, sky subtraction, wavelength 
and flux calibration), as described in \citet{fpb02}. Cosmic ray removal was 
performed using the \reduceme~package \citep{cardiel99}.  

\section{Data analysis}
\label{Sec:analysis}

\subsection{Stellar Templates}
\label{Sub:Templates}

\begin{figure}[!h]
\hspace{0.5cm}
\resizebox{\hsize}{!}{\includegraphics[angle=90]{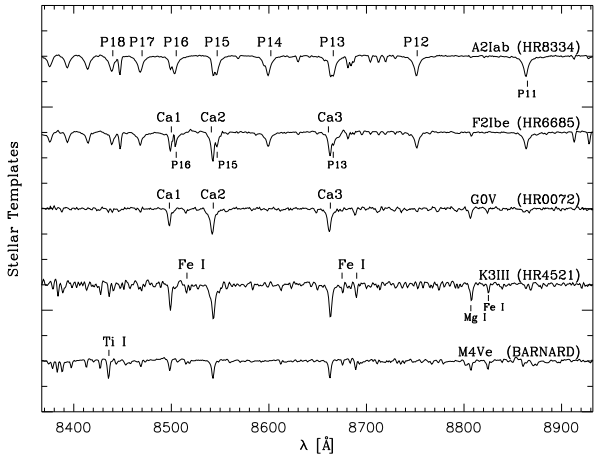}}
\caption{Representative sample of continuum subtracted stellar templates 
used in this paper. The most important features found in those spectra have 
been indicated: the Paschen series (from P11 to P18), the Ca {\tiny II} 
triplet (Ca1, Ca2 and Ca3) and several metal lines (Fe\,{\sc i}, Mg\,{\sc i}
and Ti\,{\sc i}).}
\label{Fig:Stemplates}
\end{figure}

In Table~\ref{Tab:templates} we list the observed stellar templates. A
representative set of spectra is shown in Fig.~\ref{Fig:Stemplates}. The wide
range of spectral types covered allows us for an exploration of the best
templates for matching galaxy spectra. The most apparent variation from top to
bottom in Fig.~\ref{Fig:Stemplates} is the  decrease of the Paschen lines
toward later types.  P14, in a region devoid of  other features, is a strong
discriminant of the presence of stars earlier  than G. P13 and P15, when
present, give characteristic asymmetric profiles  to the second and third lines
of the Ca\,{\sc ii}  \citep{cenarro01b}. 
Following previous works in the literature \citep{rw92,hau99} we have 
included in our analysis a mix-template from a linear combination of stars in 
our sample. However we have used a different approach to derive our optimal 
mix-template. Given the importance of the continuum shape in this wavelength 
region \citep{cenarro01, vazdekis03}, we have created an optimal template that provides 
the best fit to the overall continuum shape of each galaxy (i.e. fitting the 
regions outside the main Ca\,{\sc ii} features, but including the TiO bands), 
before broadening with the appropriate kinematics. It is important to perform this fit 
using a combination of unconvolved stellar spectra, since any input kinematics comes 
from a particular template and could bias our result. In most cases the optimal 
mix-template is formed by G8III, K3Ib and M4V stars. The importance of the M4V star in the
final template is justified by its ability to match the TiO molecular bands. As we will 
show in Sect.~\ref{Sub:Bestfit}, this mixture of stellar templates helps to alleviate the 
template mismatch problem and it is an ideal benchmark to test our SSP models. 

\begin{table}[!h]
\begin{center}
\caption{\sc Stellar Templates}
\label{Tab:templates}
{\tabcolsep=3pt
\begin{tabular}{ccccc}
\hline\hline
  Name  & $\alpha$ (h m s)& $\delta$ (d m s) & Spectral Type & M$_V$ \\
  (1)  &  (2)  & (3)  & (4)  & (5)  \\
\hline
BARNARD$^1$ & 17 57 48.50 & $+$04 41 36.2 & M4 Ve   & 9.54\\
 GL818$^2$  & 21 05 19.75 & $+$07 04 09.5 & K6 V    & 8.33\\
 HR0072     & 00 18 41.87 & $-$08 03 10.8 & G0 V    & 6.46\\
 HR4521     & 11 46 55.62 & $+$55 37 41.5 & K3 III  & 5.26\\
 HR5340     & 14 15 39.67 & $+$19 10 56.7 & K3 Ib   & 1.23\\
 HR5681     & 15 15 30.16 & $+$33 18 53.4 & G8 III  & 3.47\\
 HR5826     & 15 31 24.93 & $+$77 20 57.7 & K5 III  & 5.00\\
 HR5854     & 15 44 16.07 & $+$06 25 32.3 & K2 III  & 2.64\\
 HR6685     & 17 55 25.19 & $+$26 03 00.0 & F2 Ibe  & 5.47\\
 HR8334     & 21 45 26.93 & $+$61 07 14.9 & A2 Iab  & 4.31\\
 HR8795     & 23 07 00.26 & $+$09 24 34.2 & M1 III  & 4.55\\
\hline\hline		    
\end{tabular}}
\end{center}
{\small {\bf \sc Notes:} 
\newline (1) HR names from the Bright Star Catalogue \citep{bsc}.
\newline (2),(3) J2000 coordinates from SIMBAD
\newline (4), spectral type from \citet{worthey94}
\newline (5) from SIMBAD (http://simbad.u-strasbg.fr/Simbad)
\newline $^1$ NSV~9910 from the New catalogue of suspected variable stars \citep{ncsv}
\newline $^2$ HD~200779 from the Henry Draper Catalogue and Extension \citep{hdcat}}
\end{table}

\subsection{SSP model templates}
\label{Sub:SSPmodels}

We have used 78 SSP synthetic models at FWHM=1.5~\AA\ (0.85~\AA/pix) in  the
Ca\,{\sc ii} triplet region, corresponding to a single-age, single-metallicity
stellar populations of [Fe/H]=$-$0.7,$-$0.4, 0.0, and $+$0.2. For each
metallicity, the age varies from 1 to 17~Gyr, except for [Fe/H]=$-$0.7 starting
from 6~Gyr and for [Fe/H]=$-$0.4 starting 3~Gyr. These limitations arise from the
limited coverage of atmospheric parameters of the stellar sample used by the
models. More details can be found in \citet{vazdekis03}. In order to adapt these models
to the instrumental resolution of our data, we have convolved them with a
Gaussian of $\sigma$=0.38~\AA. This procedure relies on the assumption that a
Gaussian represents a good match to the instrumental PSF (Point Spread Function)
of the observations. We have tested whether the arclines followed a gaussian
shape, finding that indeed they do, with only small deviations (below 0.01) in
the $h_3$ and $h_4$ parameters, as described by \citet{vdm93}. If $h_3$ and
$h_4$ had been found to be significantly different from 0, we would have had
to  convolve the models with the line profile of the arc-lines.

\subsection{Extraction of the kinematics}
\label{Sec:Kinextract}

Kinematic profiles were derived from the data in the region of the near-IR 
Ca\,{\sc ii} triplet (8498~\AA, 8542~\AA, 8662~\AA).  For each galaxy, we 
generated  two sets of coadded galaxy spectra, with minimum signal-to-noise ratios 
(S/N) of 33 and 20 per pixel, respectively.  We made use of the program FOURFIT 
developed by \citet{vdm93}, available from the first author's website. A  Gauss-Hermite 
fit to the LOSVD was performed to obtain mean radial velocities,  velocity dispersions, 
$h_3$ and $h_4$ moments for the data with signal-to-noise ratio above 33 per pix$^{-1}$. 
For data with signal-to-noise ratio between 33 and 20 per pix$^{-1}$ a pure gaussian fit 
was used. The choice of S/N is explained in Sect.~\ref{Sub:Errsim}.

\subsection{Selection of the best fitting template}
\label{Sub:Bestfit}

We have determined the kinematic profiles for our sample using each of the 90
templates we had available (78 SSP models, 11 stellar templates and the best
mix-template for each galaxy). We have considered as our best fit template the 
one that minimized the $\chi^{2}$, as defined in \citet{vdm93}, for the central 
position of the galaxy profile, which has a S/N well above 33. One of the features 
of this code is the possibility of filtering the input spectra in Fourier space 
so that any remaining low-order component not  removed by the continuum subtraction 
or high-frequency noise can be filtered out. The resulting $\chi^{2}$ is then 
evaluated over a finite range of wavenumbers determined by the filtering parameters. 
For our  dataset the results from FOURFIT were weakly dependent on the lower cutoff 
and insensitive to the choice of the upper cutoff (given that we chose the same upper 
cutoff for all the galaxies and that this cut was well above the sigma of the galaxy). 
We also reduce the uncertainties in the derived kinematics by flux-calibrating the 
data and the template stars before applying the continuum subtraction because the 
normalizing polynomia for the data should be very similar to the ones used for the 
templates. otherwise, the normalization procedure might introduce non-negligible 
variations in the spectra, including in the equivalent widths of the lines.

\begin{figure}
\resizebox{\hsize}{!}{\includegraphics[angle=0]{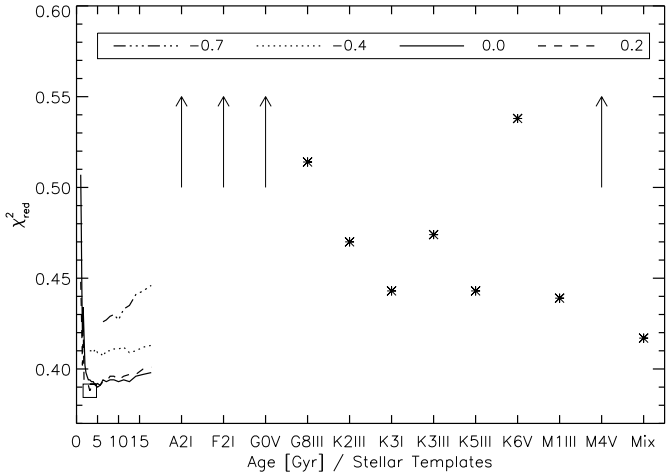}}
\caption{$\chi^2$ distribution for NGC~5326. Open square indicates the 
minimum $\chi^2$ for the best SSP fitting model, from which kinematics has been 
derived for this galaxy. $\chi^2$ values for the stellar templates 
A2 Iab, F2 Ibe, G0 V and M4 Ve have been plotted with arrows indicating that they 
fall outside the plot limits.  Their $\chi^2$ values are: 2.197, 1.313, 0.613, 
1.715 respectively.}
\label{Fig:Chi2}
\end{figure}

We establish whether SSP, stellar templates or mix-template provide better
fits to the galaxy spectra by comparing the global reduced $\chi^{2}$ of the
fits.  A typical distribution of $\chi^{2}$ values is shown in 
Fig.~\ref{Fig:Chi2}.  The lines trace reduced $\chi^{2}$ for the SSP  models,
labeled by age and metallicity, while the crosses are the reduced  $\chi^{2}$ of
the stellar templates labeled by their spectral types and combined stellar
template.  The patterns in this Fig. are representative of those found for
all the  galaxies in the sample.  There is a group of stellar templates that do 
significantly better than the others; in particular, the A2, F2 and M4 
templates consistently yield poor fits in all the galaxies.  For the SSP 
models, the $\chi^{2}$ distributions are fairly flat as a function of age for a
given metallicity, except for ages below $\sim$2~Gyr, where the quality  of the
fits quickly degrades.  This behavior is expected since, at younger ages,
the Ca\,{\sc ii} equivalent widths significantly decrease and the Paschen series
become noticeable (see \citealt{vazdekis03}).  At each age, the fits are sensitive to 
metallicity.  

For the spectrum used in Fig.~\ref{Fig:Chi2}, the best fit is obtained with a
SSP model corresponding to [Fe/H]=+0.2 and 3.16~Gyr. It is worth noting that
this SSP model should not necessarily be considered as the best fit to the
stellar populations of this galaxy (see below). For the entire sample, the 
$\chi^{2}$ for the best SSP model, best stellar template and mix-template are 
quite similar, and for 16 of the 19 objects the SSP templates provide marginally 
better fits. This result demonstrates that synthetic SSP spectra, convolved to 
the resolution of the data, can provide accurate templates for kinematic analysis 
of galaxy spectra. The high quality of the SSP fits is most-likely due to their 
ability to minimize the template mismatch problem. While the stellar populations 
of bulges are likely to contain a mixture of ages and a range of metallicities, the 
SSP models provide a better approximation to galaxy spectra than a single stellar
spectrum.  Our results indicate that late KIII stars and early M stars provide
a better approximation to the integrated  light of bulges than A, F, and G stars
in agreement with  \citet{vazdekis03}. Table~\ref{Tab:Sample}, Col.~13, lists the 
overall best template for each galaxy. Note that, for three galaxies (NGC~5389, 
NGC~5719, NGC~5746), the M1III and K3I stellar templates provide the overall 
best fit to the galaxy spectra. 

\begin{figure*}[!t]
\includegraphics{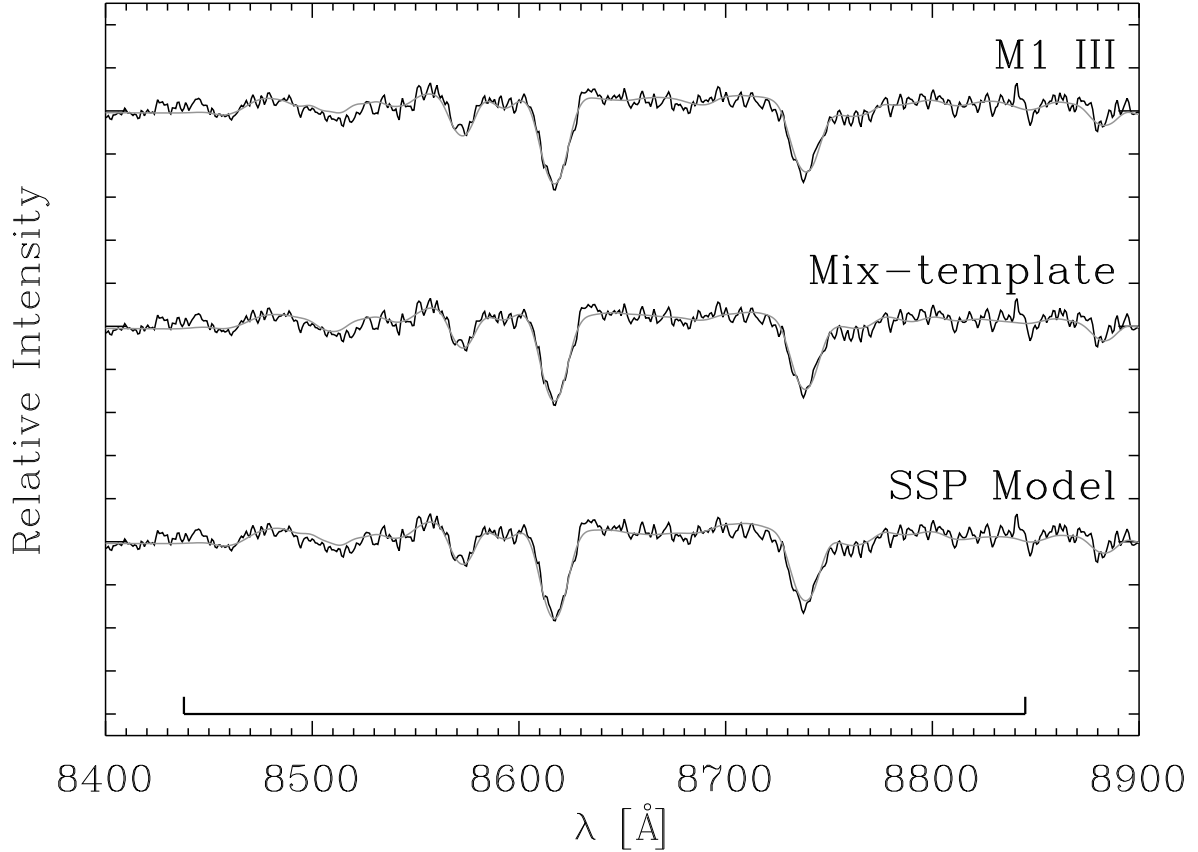}
\caption{Comparison between our best stellar template, mix-template, 
and best SSP model fits, for the continuum subtracted and low-order 
filtered, binned central spectrum of NGC~5326. The fitting region 
has been indicated with a line at the bottom of the Fig. In this 
case the best fitting template is the SSP model.}
\label{Fig:comparison}
\end{figure*}

\begin{figure}
\resizebox{\hsize}{!}{\includegraphics{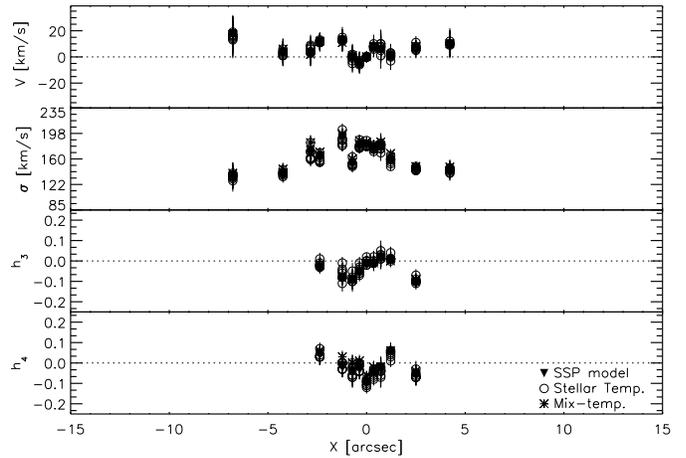}}
\caption{Comparison of the profiles obtained with our best stellar templates 
(G8III, K2III, K3I, K3III, K5III, K6V, M1III), mix-template and best SSP 
model fit of NGC~5326.}
\label{Fig:comparingprof}
\end{figure}

In Fig.~\ref{Fig:comparison} we plot the central spectrum of NGC~5326 and the
best-fitting stellar (M1 III), mix and SSP templates, convolved with their
respective kinematic solutions.  The SSP model represents a very good fit for
the whole sample of galaxies spectrum. Fitted output parameters for the best 
template stars (G8III, K2III, K3I, K3III, K5III, K6V, M1III), mix-template and 
best SSP model are shown in fig.~\ref{Fig:comparingprof} for the entire profile 
of NGC~5326.  The results show an excellent agreement. Therefore we adopt the 
solution from our best SSP template for each galaxy to obtain kinematic profiles.

A point to remark is that no inferences on the stellar population age and 
metallicity of the galaxies are presented here from the best-fitting SSP models. 
The reason is that an important relation has been shown to exist between
the continuum shape and the stellar populations in this spectral region 
\citep{cenletter,vazdekis03}. Since we remove this continuum, as required by FOURFIT, we
neglect its valuable information for constraining galaxy stellar populations. Therefore 
the most useful information left in our spectra is the equivalent width of the 
Ca\,{\sc ii} feature, which does not vary much for SSPs of the age and metallicity 
regimes considered here. The wavelength region that we have used to derive the 
kinematics excludes the strong TiO molecular bands seen in this spectral range. 
Therefore we  are left with an almost featureless continuum shape which is 
straightforward to subtract. A different methodology is required to be able to 
analyze their stellar populations on the basis of the SSP models. Methods for 
extracting useful information from the full SSP spectra are discussed in 
\citet{vazdekis99} and \cite{vazari99}. The stellar population analysis of this 
galaxy sample will be presented in a forthcoming paper \citep{fpvb03}.

The use of SSP models has a potential to minimize the template mismatch
problems that have traditionally affected galaxy kinematic measurements. 
Whereas the KIII stellar templates have been shown to provide reasonably good
fits to elliptical galaxy spectra in the optical region, one expects that the
SSP models should significantly improve the kinematical fits. The blue and
visible have been proven to have large sensitivities to the age, metallicity and
abundance ratios \citep{rose94,worthey94,vazdekis96,vazdekis99}. Our results 
suggest that the requirement of observing template stars with the same 
instrumental setting as the galaxies is not compelling. The use of SSP models 
as templates allows us to use all of the available observing time for the 
targets, without the need to take spectra of template stars. An important 
application of SSP models should be in the field of high-redshift galaxy 
kinematics, where the measurement of the same spectral range in stars and 
target galaxies would require changing the instrumental setup, and where the 
redshift of the target objects is not generally known a priori.

The main limitation of the use of SSP models is that it relies on a well 
designed stellar library to produce the full stellar population synthesis.  
However, the method is expandable to include combinations of SSP models to 
account for composite populations.  Potential applications are the study 
of age differences between bulge and disk, or population signatures of 
kinematically-decoupled cores.  

\subsection{Errors \& Simulations}
\label{Sub:Errsim}

Although FOURFIT gives the formal errors of the fit, we have performed 
simulations to establish the level of confidence the code can produce on 
(V, $\sigma$, $h_3$ and $h_4$) at S/N=33 pix$^{-1}$, and (V, $\sigma$) at 
S/N=20 pix$^{-1}$. 

For the simulations two different tests have been performed. In the first one we 
convolved an SSP model ([Fe/H]= 0.0, t= 5.01 Gyr) with a wide range of LOSVDs to 
generate model galaxy spectra. The LOSVDs included several velocities, velocity 
dispersions, h$_3$ and h$_4$ mapping a representative parameter space covered by 
our sample of galaxies (V $\in$ [800,2000], $\sigma \in$ [50,300], h$_3$ and h$_4 
\in$ [-0.1,0.1]). The inclusion of higher velocities (up to 4000 km/s, to match our 
furthest galaxy) doesn't affect our results. Eight different samples of white noise
were added to our model galaxy spectra to yield S/N ratios of 20 and 33 pix$^{-1}$. For
each one of the noise sample at the two S/N levels the full battery of LOSVDs was
recovered. Given the fact that all SSP models give very similar $\chi^{2}$ values, only 
a representative subsample of 3 SSP models covering the models parameter space 
([Fe/H]=-0.7, Age=7.98 Gyr; [Fe/H]=0.0, Age= 5 Gyr; [Fe/H]=0.2, Age=15.85 Gyr) 
was used to perform these simulations. We have simulated both Gaussian and non-Gaussian
LOSVDs, however we only fitted pure gaussians to the spectra of S/N=20 pix$^{-1}$, for 
which we only modeled the V and $\sigma$ errors due to LOSVD mismatching in the S/N=20 
spectra. In the second test we created the galaxy spectra from a stellar template (K3I). 
Spectra from the K3I, K5III and M1III stars were used as kinematic templates. The choice 
of these templates is due to their ability to produce, overall, the lowest values of 
$\chi^{2}$ when fitted to the sample of galaxies. We have convolved the model galaxy 
spectra with the same LOSVDs and added the same white noise samples as in the first test. 

\begin{table}
\begin{center}
\caption{\sc Errors from simulations}
\label{Tab:Simulations}
{\tabcolsep=2.5pt\begin{tabular}{l|ccccccccccc}
\hline
\hline
Param. & \multicolumn{2}{c}{V [km s$^{-1}$]} & ~ & \multicolumn{2}{c}{$\sigma$ [km s$^{-1}$]} & ~ &
\multicolumn{2}{c}{h$_3$} & ~ & \multicolumn{2}{c}{h$_4$}\\
\cline{2-3} \cline{5-6} \cline{8-9} \cline{11-12}     
S/N &{\small 20} & {\small 33} & ~ & {\small 20} & {\small 33} & ~ & 
{\small 20} & {\small 33} & ~ & {\small 20} & {\small 33}\\ 
std(SSP)    & 15 & 13 & ~ & 29 & 15 & ~ & -- & 0.07 & ~ & -- & 0.08\\
std(Stars)  & 18 & 16 & ~ & 34 & 18 & ~ & -- & 0.07 & ~ & -- & 0.10\\
\hline
\hline
\end{tabular}}
\end{center}
{\small {\bf \sc Notes:} 
Standard deviations for each parameter at the different signal-to-noise ratios 
measured from the simulations (see Sect.~\ref{Sub:Errsim}).}
\end{table}

The main results of the simulations are summarised in Table~\ref{Tab:Simulations}. The 
Table gives the standard deviations of the distributions (parameter$_{out}$ - parameter$_{in}$), 
obtained from the simulations in each test, for the full set of parameters and white 
noise samples. This is a measurement of the uncertainty when working at the mentioned 
S/N ratios. We use the uncertainties obtained from the simulations with the SSP models 
and display them as error bars on the right-hand side of each kinematic profile, see 
Appendix~A. In Fig.~\ref{Fig:deltaparam}, the left panels show the distribution of the 
recovered LOSVDs compared to the input ones at each sampled value of $\sigma$ and h$_4$ 
for each test for S/N$>$33. The values shown below each histogram represent the median
(within square brackets) and standard deviation of the distribution, found at those values 
of $\sigma$ and h$_4$. The right panels, on the other hand, show the total distribution 
and standard deviation (i.e. the sum of the individual ones) from which we have obtain 
the values shown in Table~\ref{Tab:Simulations}. No significant differences are found 
between the distributions obtained from the SSP models and stellar templates. This 
result strengthens our approach of using SSP models as kinematic templates.

The simulations show two effects worth mentioning. First, there is no systematic 
underestimate of the velocity dispersion for low input velocity dispersions 
(see Fig.~\ref{Fig:deltaparam}). This effect, if present, might bias our velocity 
dispersion results for our two smallest bulges, see Fig.~\ref{Fig:sigmaprof}. Such an
effect has been previously reported by \citet{joseph01} using a different code (i.e. FCQ), 
finding an overestimation of $\approx$10 km/s at $\sigma$=50 km/s. Second, as we move 
towards larger velocity dispersions, the uncertainty in $\sigma$ also increases (from 6 
km/s at $\sigma$=50 km/s, to 15 km/s between $\sigma$=100-200 km/s and 32 km/s between 
$\sigma$=250-300 km/s). Hence, although the Gauss-Hermite expansion conforms a description 
of the LOSVD with a set of orthogonal functions and therefore we only expect major 
coupling between  $\gamma$ (the peak of the LOSVD) and $\sigma$ \citep{larsen83}, V and 
$h_3$, and  $\sigma$ and $h_4$, we have detected other correlations when working at low 
signal-to-noise ratios. The expected coupling of V with $h_3$ and $\sigma$ with $h_4$ 
appears in our simulations. Similar results have been found by authors who have performed 
similar tests \citep{vdm93,hau99}. 

We have decided to impose a S/N=33 as the minimum level to determine $h_3$ and
$h_4$. This is a slightly lower level than that suggested by \citet{vdm93}
(S/N$\simeq$40 pix$^{-1}$) to give a good estimation of the Gauss-Hermite higher
order moments. Nevertheless in most cases we obtain S/N$\geq$40 for the
profile central position (see Table~\ref{Tab:Sample}).

\begin{figure}
\resizebox{\hsize}{!}{\includegraphics{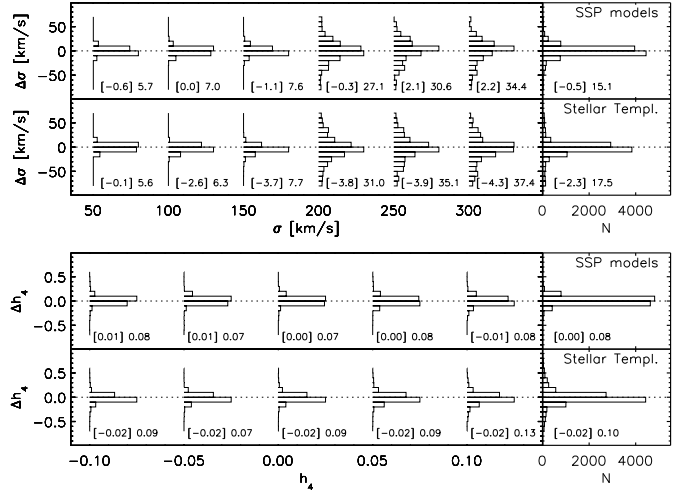}}
\caption{Recovery of $\sigma$ and h$_4$ via FOURFIT using SSP models and stellar
templates. The left panels show the distribution of the recovered LOSVDs compared 
to the input ones at each sampled value of $\sigma$ and h$_4$ for each test at  
S/N$>$33. The right panels, on the other hand, show the total distribution 
(i.e. the sum of the individual ones on the right panels). For each $\sigma$ and h$_4$
we plot a histogram of the $\Delta$($\sigma$)=$\sigma$(out) $- \sigma$(in) and 
$\Delta$(h$_4$)=h$_4$(out) $-$ h$_4$(in) obtained from the 14400 realisations 
corresponding to 8 white noise realizations. The values shown below each histogram 
represent the median (within square brackets) and standard deviation found at those 
values of $\sigma$ and h$_4$.}
\label{Fig:deltaparam}
\end{figure}

\subsection{Measurement of the CaT$^*$ index}
\label{Sec:CaT}

We have measured the recently defined Paschen-free CaT$^*$ index
\citep{cenarro01}, that  measures the strength of the Ca\,{\sc ii} lines
corrected from the contamination by Paschen lines, from our flux calibrated
spectra. We have made used of the {\small \sc FORTRAN} routine {\tt indexfits}
(part of the \reduceme~package), kindly made available to us by Javier Cenarro. 
No velocity dispersion correction and zero-point offset have been performed. 
Errors in the CaT$^*$ index include contributions from radial velocity 
uncertainties and from S/N. The first one is obtained by the CaT$^*$ index from 
100 random simulations within the velocity range [$V-\Delta V$,$V+\Delta V$]. 
The latter is determined using eq.~A37 from \citet{cenarro01}. Both terms are 
added in quadrature. The uncertainty from the S/N dominates the CaT$^*$ error.

\subsection{Comparison with the literature}

\begin{figure}[t]
\centering
\includegraphics[width=0.85\linewidth]{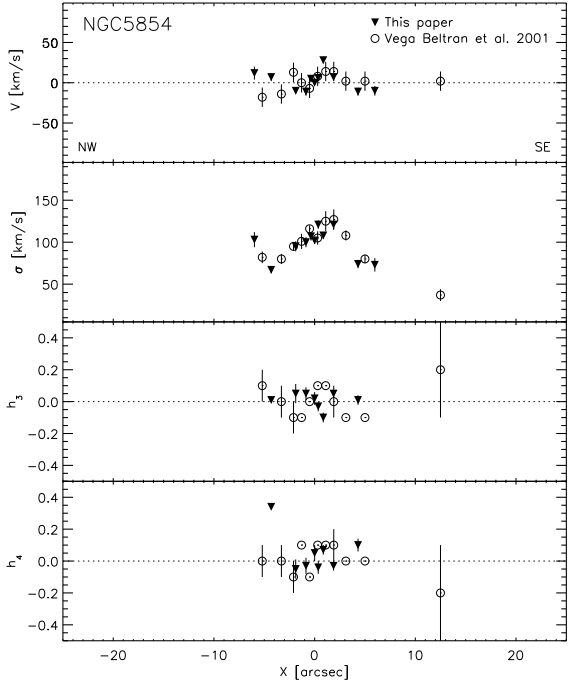}
\caption{Comparison between our best SSP model fit and \citet{JVB01} for
NGC~5854.}
\label{Fig:compN5854}
\end{figure}

\begin{figure}[t]
\centering
\includegraphics[width=0.85\linewidth]{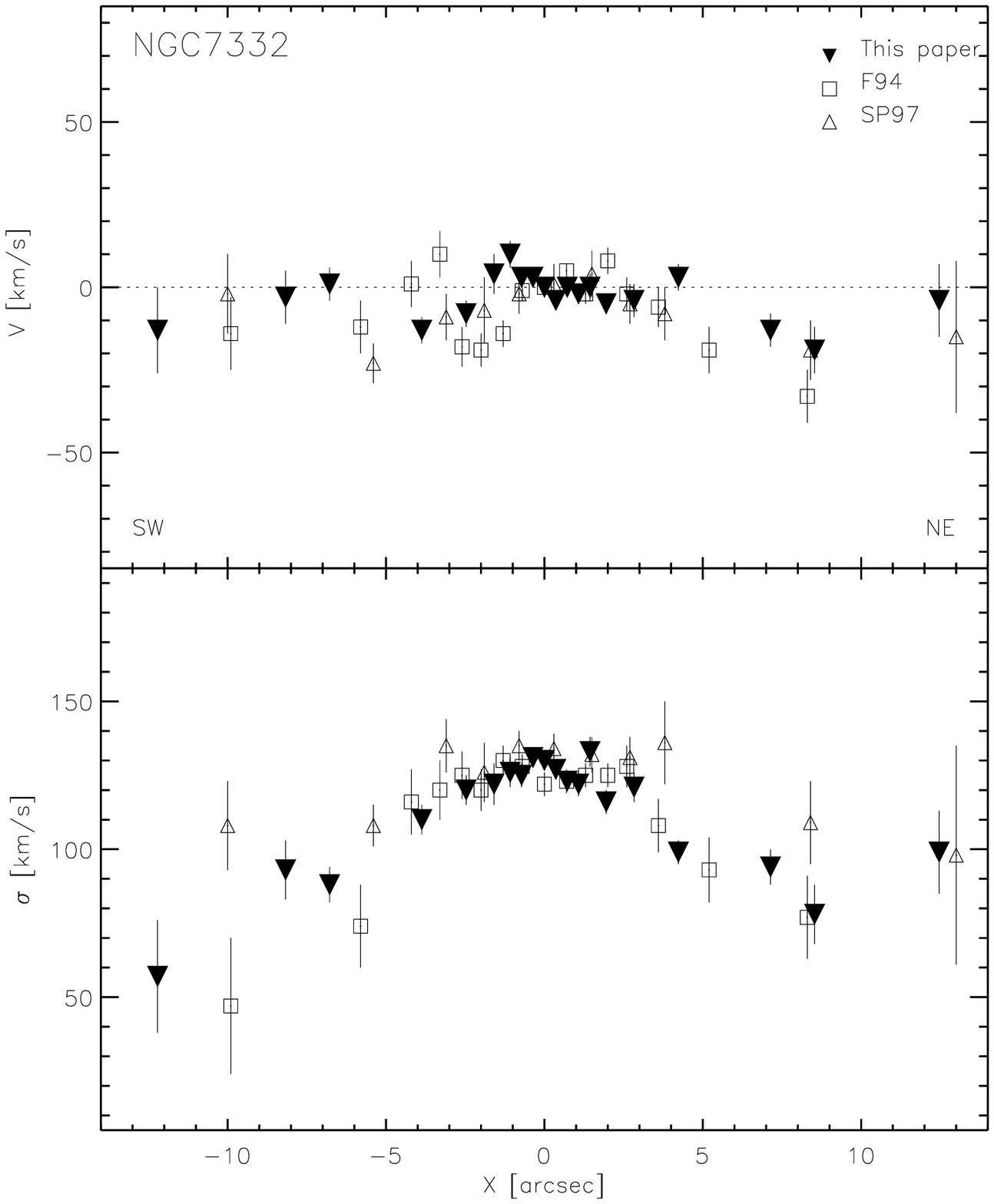}
\caption{Comparison between our best SSP model fit with \citet{f94} (F94) and
\citet{sp97} (SP97) for NGC~7332.}
\label{Fig:compN7332}
\end{figure}

We have compared our best SSP model fits with those in the literature. 
Unfortunately, published minor axis kinematic profiles are scarce.  We 
present the comparison in Fig.~\ref{Fig:compN5854} and \ref{Fig:compN7332} 
for NGC~5854 and NGC~7332 respectively.  An excellent agreement is found for 
the two galaxies, with no major systematic deviations, supporting our approach 
of using synthetic SSP models.  This agreement is particularly good for the 
velocity dispersion of NGC~5854, where our results follow the same trend 
obtained by \citet{JVB01}. The comparison for NGC~7332 reveals also good 
agreement with \citet{f94} and \citet{sp97}.

\section{Results}
\label{Sec:Results}

Kinematic profiles for the 19 galaxies in the sample are shown in
Fig.~A.1 (Appendix A). For each galaxy we show profiles of mean
heliocentric velocity $V$, velocity dispersion ($\sigma$), $h_{3}$ and $h_{4}$
Gauss-Hermite coefficients, and Ca\,{\sc ii} triplet index CaT$^{*}$, along 
the galaxy's minor axis. Abscissae are centered on the brightness peak 
of the long-slit spectrum, i.e. on the photometric center of the galaxy 
since we don't expect large centering errors. Negative abscissae correspond 
to the minor axis that projects away from the disk (the ``dust-free'' side), 
whose position angle on the sky is given at the top next to the galaxy's name.
Circles are measurements derived from $S/N > 33$ pix$^{-1}$ spectra, and triangles 
are $V$, $\sigma$ and CaT$^{*}$ derived from $20 < S/N < 33$ spectra. Error bars 
in each point are the formal error estimates coming from FOURFIT code.
Measurement uncertainties derived from the simulations described in 
Sect.~\ref{Sub:Errsim} are shown with vertical bars on the right edge of each 
plot for the $S/N > 33$ spectra and, for $V$ and $\sigma$, for $20 < S/N < 33$ as well 
(longest bar).  The contour plot draws isophotes of the HST NICMOS F160W images
\citep{pb99}, drawn to scale. The length of the 1.2 arcsec wide slit delimits the 
range along the minor axis profile from which spectra has been binned to obtain a 
S/N$>$20. The high spatial resolution of the NICMOS images (0.075 arcsec/pixel, 
PSF=0.18 arcsec FWHM) provides a clue on the level of substructure existing within the 
spectroscopic slit. Due to the lack of a NICMOS image for NGC~7332, a WFPC1 image, retrieved 
from the HST archive, has been used instead. Vertical dotted lines indicate the minor axis 
bulge radius and center of the galaxy (see Table~\ref{Tab:Sample} for more details). 

Owing to the short integration times and the S/N requirements for kinematic 
extractions derived from the simulations (Sect.~\ref{Sub:Errsim}), the kinematic
profiles typically extend to 5-10 arcsec from the nucleus only.  Only for the 
large galaxy NGC~7331 do kinematic profiles extend over 20 arcsec on the dust-free side.  
All profiles are drawn to the same horizontal scale. In the following subsections 
we briefly address the global behavior of each kinematic parameter.  

\subsection{Velocity profiles}
\label{Sec:Velocities}

As expected for the minor-axis alignment of the slits, mean velocity 
profiles are fairly flat.  However, the profiles are not entirely 
featureless.  We note three types of features:

\subsubsection{Rotation in the outer parts}

We detect non-zero rotation in the outer parts of the profiles
(opposite mean velocities at both sides of the center, at the
2-$\sigma$ level) in 6 out of the 19 galaxies (NGC~5389, 5689, 5746,
5838, 5854, 7457).  Typical rotation amplitudes are in the range $10 <
V < 20$ km\,s$^{-1}$.  Such estimates of velocity amplitude and frequency 
of occurrence are highly dependent on the depth of the data.  For two
galaxies (NGC~5475, 5587) the profiles do not reach 5 arcsec and
contain no information on rotation at large radii.  Brighter galaxies
in which the spectra extend farthest from the center show the highest
$V$ at large radii.  In most of these cases, the outer velocity
measurements sample disk or bulge-disk transition regions.  The
non-zero velocities indicate that our spectroscopic slits were not
perfectly aligned with the line of nodes.  The slit was placed
orthogonal to the major axis of the outer isophotes.  Hence the
misalignment might reflect an isophote twist in the outer parts
sampled by the spectra, due eg.  to warps or spiral arms, or a 
distortion of the inner disk velocity field due to a bar or a spiral 
arm. Note that a misalignment of $5^\circ$ or more could easily explain 
such amplitudes in velocity. 

\subsubsection{Inner rotation}

We detect inner minor-axis rotation in 8 galaxies (NGC~5443, 5587,
5689, 5707, 5838, 5965, 6010, 7457) and possibly also in NGC~5389,
5422, 5854.  Whenever the profiles extend beyond the nuclear region,
inner rotation is delimited by a change in the velocity profile slope. 
These bulges thus contain kinematically distinct regions (hereafter
KDR).  We do not intend to set a link between these KDR and
counterrotating or otherwise kinematically decoupled cores in
ellipticals and bulges; full counterrotation cannot be established
from our minor-axis kinematic profiles.  Rather, we associate the
velocity features to slight variations in the kinematic structure,
related to inner components such as disks or bars.

Theoretically the simplest way to explain a minor-axis rotation in a
bulge is to assume we are dealing with a triaxial object.  In a triaxial
object the position angle of the apparent minor axis can be different from
that of the projected intrinsic short axis of the object
\citep{tim91}.  Moreover, the photometric and kinematic minor axes may
be misaligned as a result of the wide range of possible locations of
the total angular momentum in a triaxial object \citep{binney85}.  In
practice, whether minor-axis rotation traces a triaxial bulge can only be
checked using integral field spectrographs e.g. {\it SAURON} \citep{bacon01}, 
or measuring kinematics along different positions angles of the galaxy 
\citep{ss99}. 

The typical minor-axis radial extent of the KDR is 2--3 arcsec, or 0.5 kpc
at the median distance for the sample.  In all of the eight galaxies listed
above as KDR, isophotal analysis of the HST/NICMOS F160W images (our
work, unpublished) reveals an inflection in the $C_4$ isophotal
coefficient of the $\cos(4\theta)$ term that measures disky deviations
of the isophotes from pure ellipses (e.g. \citealt{jfk92}). $C_4$ amplitudes 
of the features are above 1\% in all cases.  This suggests the presence of
flattened subcomponents in the inner regions of the bulges showing up as 
excesses in the bulge surface brightness profiles \citep{balcells03}.  One 
probably sees a mixture of inner disks and inner bars. In four of the eight 
cases the isophotal analysis reveals changes in isophote position angle 
($\Delta$PA $\ge$ 10$^\circ$) at the same radius, which are likely to correspond to
inner bars but could also trace misaligned disks. The small rotation amplitudes
and the lack of isophotes elongated perpendicularly to the disks suggests that none of our
bulges show the type of orthogonal bulge kinematics found in NGC~4698 \citep{bertola99} 
and NGC~4672 \citep{sarzi00}. The four remaining cases with no isophotal misalignment 
probably correspond to inner bars, as an aligned disk has its line of nodes parallel to 
the minor axis.

The most prominent nuclear KDR is that of NGC~5838.  The nucleus of
NGC~5838 harbors an ordered ring or disk of dust and young stars
\citep{pb99}, which is oriented orthogonal to the slit. 
The dust may be responsible for the asymmetric rotation pattern.

It is worth noting that none of the KDR have distinct colors $B-I$ 
and $I-H$ \citep{pb99} on the dust-free side of the 
galaxies, nor do they have corresponding features in the CaT$^{*}$ 
profiles (Fig.~A1).  This may suggests that the velocity features are 
not associated to distinct stellar populations (e.g. \citealt{davies01}).  

\subsubsection{Kinematically offset nuclei}

In two galaxies (NGC~5707, 7331) the velocities at large radii appear
offset in the same direction relative to the nuclear velocity. 
NGC~5707 has a nuclear disk or bar extending over 1 arcsec, which is well
aligned with the major axis of the outer disk and partially obscured
by a dust lane; the dust might be responsible for the offset velocity
measurement at the photometric center of the galaxy.  In NGC~7331, the
entire region from -4 arcsec to +4 arcsec appears displaced 15-20
km\,s$^{-1}$ from the velocities further out.  It is unclear whether dust
extinction can be responsible for this structure, or whether the bulge
might actually wander in the potential of the galaxy. The major-axis
kinematics of this galaxy is complex, and its outer spiral structure is 
asymmetric; this could indicate a recent accretion event and/or the motion 
of the luminous material within the dark matter halo of the galaxy \citep{noor2001}. 
Results on the major-axis kinematics of this galaxy are puzzling. While \citet{prada96} 
reported to have found the bulge in NGC~7331 to counter-rotate, a more recent work 
by \citet{bottema99} contradicts this result. 
 
\subsection{Velocity dispersion profiles}
\label{Sec:Dispersions}

Velocity dispersion profiles trace the dynamical temperature of the
bulges and are key ingredients for dynamical modeling of the bulges.
They also provide diagnostics for distinguishing between bulge and disk
on the basis of the internal dynamics.

For our galaxies, the minor-axis velocity dispersion profiles show a
wide diversity of shapes, from flat to strongly centrally peaked.
Velocity dispersion gradients, ($\Delta\sigma$ per unit minor-axis
$r_e$) were computed by fitting linear relations to the
negative-abscissa dispersion profiles of Fig.~A.1, which correspond
to the dust-free side of the bulges.  Minor-axis $r_e$ were derived
using the $r_e$ values from \citet{balcells03}, which derive from
$H$-band surface brightness profiles from $HST$-NICMOS images, and
ellipticities at the bulge effective radius $\epsilon_B(r_e)$ obtained from
the ellipticity profiles in the same paper.  Dispersion gradients range
from 0 to -70 km/s/$r_e$, with a median of -21 km/s/$r_e$ and a mode of
only $\sim$10 km/s/$r_e$.

\begin{figure}[!t]
\resizebox{\hsize}{!}{\includegraphics[angle=90]{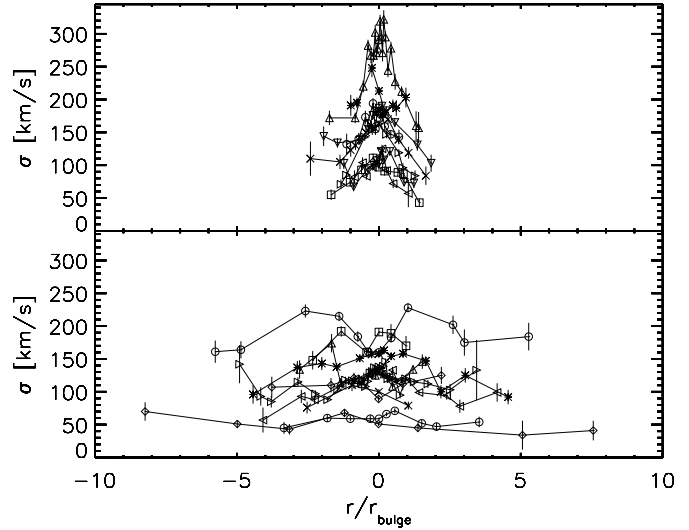}}
\caption{Velocity dispersion profiles for our sample of galaxies.
Bulges with centrally-peaked velocity dispersions profiles are
drawn in the top panel, whereas bulges with flat velocity dispersion 
profiles are plotted in the bottom panel. In the abscissa, radii have been
normalized to the minor axis bulge radius (Table~\ref{Tab:Sample}, Col.
12), defined as the radius along the minor axis where bulge and disk
have equal surface brightness.}
\label{Fig:sigmaprof}
\end{figure}

Fig.~\ref{Fig:sigmaprof} displays all the velocity dispersion
profiles, presented on two panels for clarity; the top/bottom panel
shows the half sample with highest/lowest $\nabla(\sigma)$.  In order
to show whether the profiles sample only the bulge or extend to the
disk region, we normalize each abscissa to $r_{bulge}$, the minor-axis
radius where bulge and disk contribute equally to the surface
brightness profile (listed in Col. 12 of Table~\ref{Tab:Sample}).  Most
profiles extend well beyond $r/r_{bulge}=1$, indicating that the
profiles reach to the region dominated by disk light according to the
bulge-disk decomposition.

Three effects may influence the shape of the minor axis velocity dispersion
profiles in our galaxies, (i) the intrinsic velocity structure of the
bulge, (ii) the galaxy inclination, which determines which projection
of the velocity ellipsoid is sampled by the data, and (iii) the
brightness of the disk in the region sampled by the kinematic
profiles.  

Because most profiles reach the region where the surface brightness is
dominated by the disk (Fig.~\ref{Fig:sigmaprof}), flat dispersion
profiles imply similar dispersions for bulge and host disk.  At the
median inclination of the sample, $i=73^\circ$, minor-axis dispersion
profiles preferentially sample the radial component of the velocity
dispersion ellipsoid.  Our flat profiles then indicate a surprisingly
high {\it radial} velocity dispersions in the regions dominated by disk
light, and an equally surprising continuity of the radial velocity
dispersions into the region of the bulge.  Steep dispersion profiles,
in contrast, indicate distinct radial dispersions in the bulge and host
disk, with dispersions increasing toward the center as is expected in
dynamically hot, self-gravitating ellipsoids.

\begin{figure*}[!t]
\resizebox{\hsize}{!}{\includegraphics{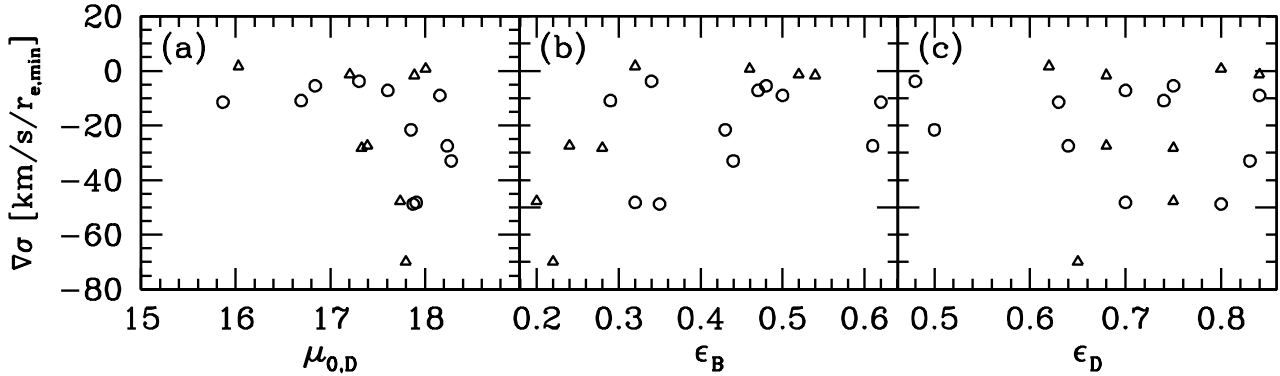}}
\caption{Dependency of the velocity dispersion gradient on galaxy parameters.  
(a) $\nabla\sigma$ against face-on $H$-band disk central surface brightness 
$\mu_{0,D}$.  (b)  $\nabla\sigma$ against bulge ellipticity $\epsilon_B$. (c)  
$\nabla\sigma$ against disk ellipticity $\epsilon_D$.  Barred or box-peanut 
shaped galaxies in our sample are represented with open triangles.}
\label{Fig:sigmagrad}
\end{figure*}

The disk central surface brightness influences the velocity dispersion
gradient.  Fig.~\ref{Fig:sigmagrad}a shows the distribution of
$\nabla(\sigma)$ against the face-on disk central surface brightness
$\mu_0$.  Bright disks have shallow dispersion profiles, while the
steepest profiles are found in faint disks.  We find it unlikely that
this dependency arises from biases in the derivation of the kinematic
profiles.  Our simulations (Sect.~\ref{Sub:Errsim}) show that
dispersions as low as $\sigma \sim 50$ km/s are well reproduced by the
FOURFIT algorithm.  Selection biases would hamper the detection of disk
dispersions for faint disks first, but dispersion gradients of all
values are found for these disks.  The lack of steep dispersion
profiles in bright disks appears therefore real.

Fig.~\ref{Fig:sigmagrad}b shows the distribution of $\nabla(\sigma)$
against $\epsilon_B(r_e)$, the ellipticity measured at the bulge
effective radius.  A trend is found, in the sense that more flattened
bulges show shallower dispersion profiles.  This dependence does not
arise from viewing-angle effects: Fig.~\ref{Fig:sigmagrad}c shows that
$\nabla(\sigma)$ does not depend on the galaxy inclination.  
Shallow-profile objects may be kinematically described as thickened disks. 
In low-ellipticity bulges, the $\sigma$ profiles increase toward the
center, again as expected from hot ellipsoids.

\citet{kor93} discusses that some bulges share strong similarities with
disks, on the basis of their low velocity dispersions, high $V_{\rm
max}/\sigma_0$ for their ellipticity, and nuclear spiral structure.
The disky nature of bulges is usually discussed in relation to
late-type, low-mass bulges \citep{kor93}. Our dispersion profiles indicate
that this situation affects also early-types, including several S0's.
  
It appears that the disky nature of bulges cannot be established on the basis 
of spheroid luminosity; our velocity dispersion gradients do not correlate 
with bulge luminosity or with central velocity dispersion.  Gradients 
do not correlate either with morphological type index, bulge S\'ersic 
index $n$, bulge and disk scale lengths and bulge effective surface 
brightness.  We have searched for a dependency on the presence of bars 
or box-peanut shaped bulge isophotes.  Such galaxies are depicted with 
triangles in Fig.~\ref{Fig:sigmagrad}.  While bar detection is difficult 
at the inclinations of our sample galaxies, it is apparent that barred 
and box-peanut bulges do not occupy preferred regions in 
Fig.~\ref{Fig:sigmagrad}.

The massive bulge of NGC~5746 (Sb) is the only object showing a
strongly-defined central velocity dispersion minimum. A weak central 
minimum is also found in NGC~5443.  Central velocity dispersion minima 
are generally understood as tracing embedded disks, bars, or low-mass spheroids.  
Therefore, given the abundance of isophotal substructure in the nuclei 
of these galaxies, it is surprising that central velocity dispersion minima are 
so infrequent among our objects.  As mentioned above, our sample includes six 
box- or peanut-shaped bulges.  For these objects, the bar producing the peanut
may not reach the center; or, alternatively, the central dispersion may
be affected by the presence of secondary {\sl nuclear} stellar bars/disks,
with dispersion minima in bars being found for given viewing angles only.  In
NGC~5746 the local central minimum may be due to the main bar reaching
all the way to the inner arcsecond, or to the presence of a secondary
bar or nuclear disk.

\subsection{High-order Gauss-Hermite terms}
\label{Sec:h3h4}

Profiles of the $h_{3}$ skewness term of the LOSVD are consistent with
zero in most cases.  In the few cases with significant non-zero
$h_{3}$ (NGC~5475, 6010, 7331), $h_{3}$ does not have an opposite sign
to the mean velocity that is typical of major axis profiles.  This is 
reasonable as any rotation detected in the spectra is due to 
slight slit misalignment respect to the minor axis.

Similarly, the kurtosis profiles $h_{4}$ are generally flat and close 
to zero.  In 12 galaxies however, $h_{4}$ is significantly positive 
above 0.1 somewhere in the profile, denoting centrally-peaked LOSVDs.

\subsection{The CaT$^*$ profiles}
\label{Sec:CaTprofiles}

We have measured the CaT$^*$ profiles for our sample of galaxies in order to 
trace correlations between kinematic and population features along the galaxies 
minor axis. Our analysis of the profiles does not show any correlation with either
kinematic (V, $\sigma$, h$_3$, h$_4$) or photometric parameters (from
\citealt{balcells03}). The lack of correlation might be due to the very small 
dependence of the CaT$^*$ on age and metallicity for the values we expect for 
this type of galaxies \citep{vazdekis03}.

In a forthcoming paper \citep{fpvb03} we will present an extensive study of 
the near-IR Ca\,{\sc ii} triplet - $\sigma$ relation for our sample of bulges.
In that paper we give measurements of the recently defined near-infrared 
Ca\,{\sc ii} triplet indices (CaT, CaT$^*$), Paschen (PaT) and Magnesium 
(Mg\,{\sc i}) indices (see \citealt{cenarro01}) for our sample of bulges. We show 
that both the CaT$^*$ and CaT indices decrease with central velocity dispersion 
$\sigma$  with small scatter. The decrease of CaT and CaT$^*$ with $\sigma$ contrasts 
with the well-known increase of another $\alpha$-element index, Mg$_2$, with $\sigma$. 
 
\section{Conclusions}
\label{Sec:Conclusions}

We have extracted minor-axis kinematic profiles from spectra in the Ca\,{\sc ii}
near-IR region for a well-studied sample of early - to intermediate-type,
inclined galaxies from the \citet{bp94} sample. SSP models from \citet{vazdekis03} at
1.5~\AA\ resolution have  been used in addition to the traditional template
stars. We have compared results obtained with the different templates and
determined that synthetic models produce slightly better results than those
from stellar templates, in the Ca\,{\sc ii} triplet region. We discuss the
advantages/disadvantages of using SSP models instead of stellar templates. Our
main conclusion is that synthetic models, when coming from a well constructed
stellar library, help to reduce several drawbacks such as the template mismatch
problem affecting all the traditional kinematical methods. We find a
good agreement with high-quality kinematic profiles from the literature. 
SSP model spectra used as templates have clear applications for kinematic
studies of galaxies at high redshift. 

We find  minor axis rotation in almost half of our galaxy sample. We discuss two
types of minor axis rotation: rotation in the outer parts, probably due to slit
misalignments, and inner rotation, which are associated to disky deviations of
the isophotes from pure ellipses. We also find some cases of kinematically
offset nuclei (NGC5707 and NGC7331) for which we do not find a clear
explanation. We show that flattened bulges tend to show shallow velocity dispersion
gradients and similar dispersions in bulge and inner disk, possibly indicating 
a thickened disk structure for these bulges.  Finally, we find that our velocity 
dispersion gradients correlate with disk central surface brightness and 
with bulge ellipticity, but do not correlate with disk ellipticity.  Gradients 
do not correlate either with bulge luminosity, bulge central velocity              
dispersion, bulge S\'ersic index $n$, bulge effective surface                  
brightness, galaxy morphological type, and disk or bulge effective              
radii.

\begin{acknowledgements}
JFB acknowledges the financial support from a PPARC studentship. The William Herschel 
Telescope is operated on the island of La Palma by the Isaac Newton Group in the 
Spanish Observatorio del Roque de los Muchachos of the Instituto de Astrof\'\i sica 
de Canarias. We acknowledge the professional help of the observatory staff in the 
operation of the WHT and the ISIS spectrograph.
\end{acknowledgements}


\appendix
\section{Profiles}
\label{Ap:ProfilePlots}
In this section we present kinematic profiles for each galaxy (heliocentric
velocity, velocity dispersion, h$_3$, h$_4$), together with an $H$-band NICMOS 
(F160W) contour map from \citet{pb99} (except for NGC~7332 for which a WFPC1 
image, taken from the archive, is plotted instead). The length of the 1.2 arcsec 
wide slit delimits the range along the minor axis profile from which spectra 
has been binned to obtain a S/N$>$20. The CaT$^*$ index (see Sec.~\ref{Sec:CaT}) 
is shown in the bottom panel. Filled circles represent data points with signal 
to noise ratio above 33, whereas filled triangles represent data points with 
signal to noise between 20 and 33. Contour levels have 0.5 mag difference. The 
left hand side of the profiles, for all galaxies, corresponds the dust-free side, 
while the right hand side corresponds to the dusty side. The PA (on top of the 
figure) is the position angle (N-E) of the dust-free minor axis. Error bars 
determined in the simulation for each S/N (see sec.~\ref{Sub:Errsim}) are located 
on the right hand side of each panel, S/N=33 (short bar) and S/N=20 (long bar). A 
dotted line has been drawn, on each panel, at the level of the central value for 
the velocity, velocity dispersion and CaT$^*$ index and it is at the zero value 
for the $h_3$ and $h_4$ parameters. Vertical dotted lines indicate the minor axis 
bulge radius and center of the galaxy (see table~\ref{Tab:Sample} for more details).

\clearpage
\begin{figure*}
\vspace{2cm}
\centering
\includegraphics[height=10.5cm,angle=0]{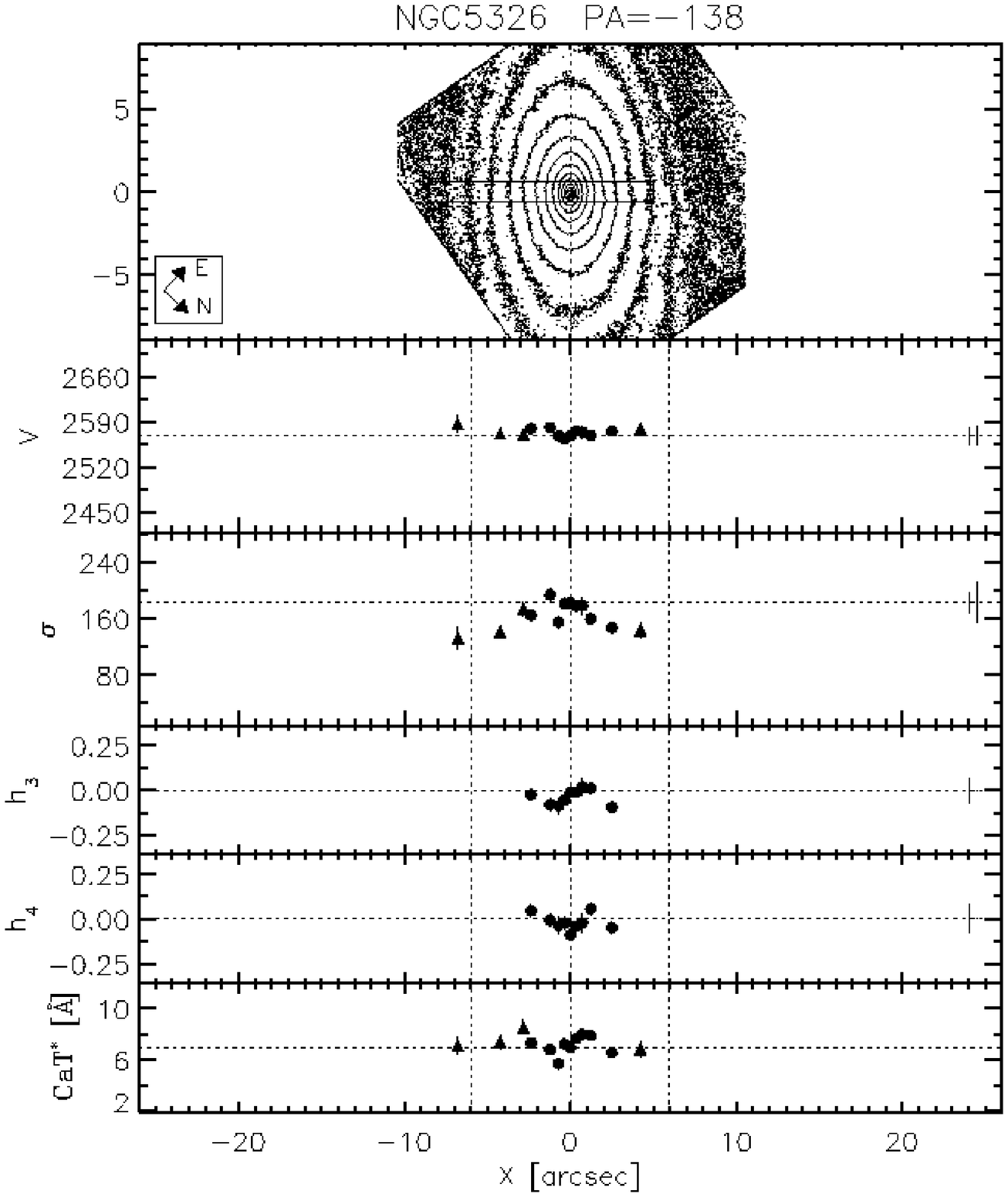}
\hfill%
\includegraphics[height=10.5cm,angle=0]{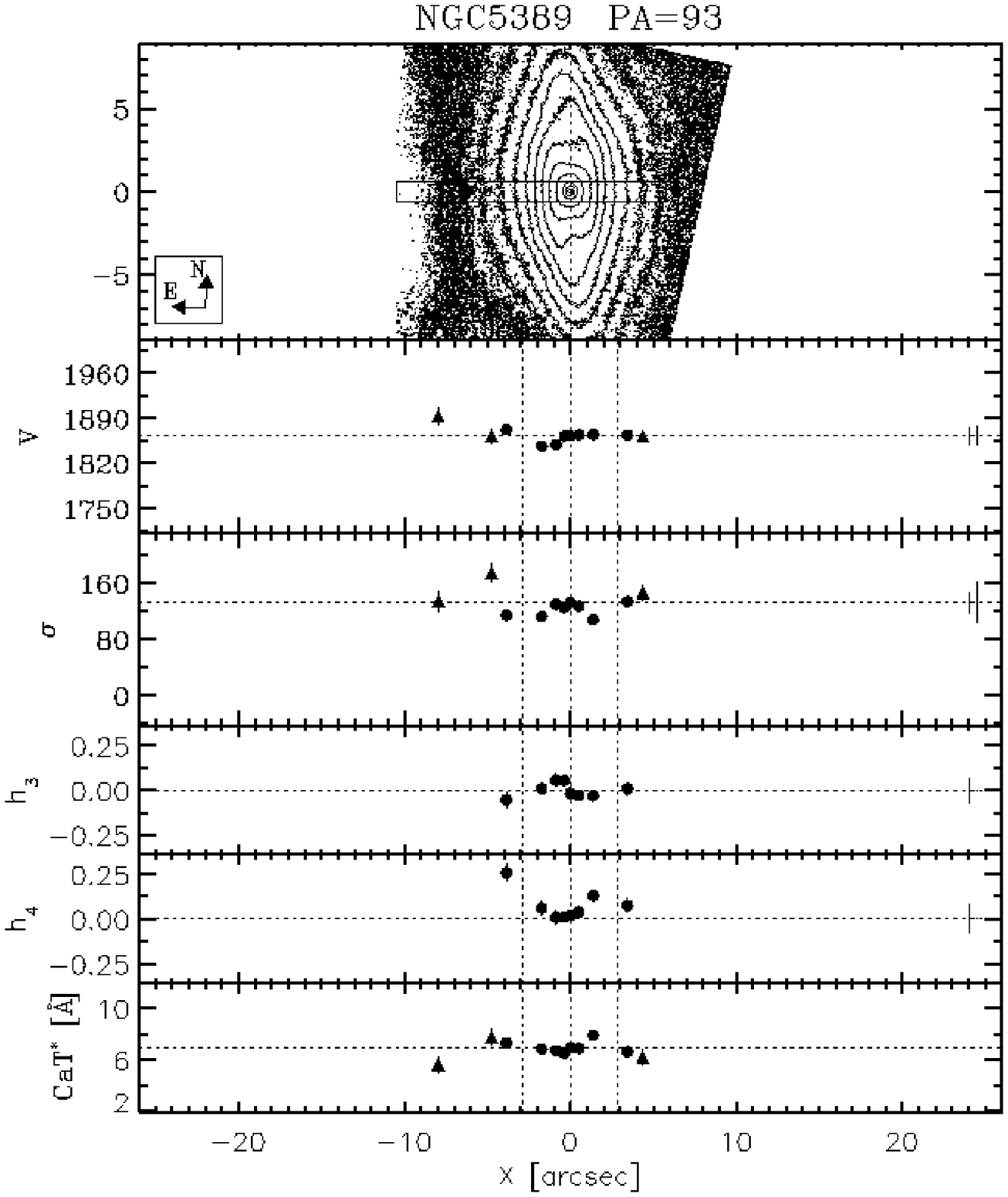}
\vspace{2cm}
\\
\centering
\includegraphics[height=10.5cm,angle=0]{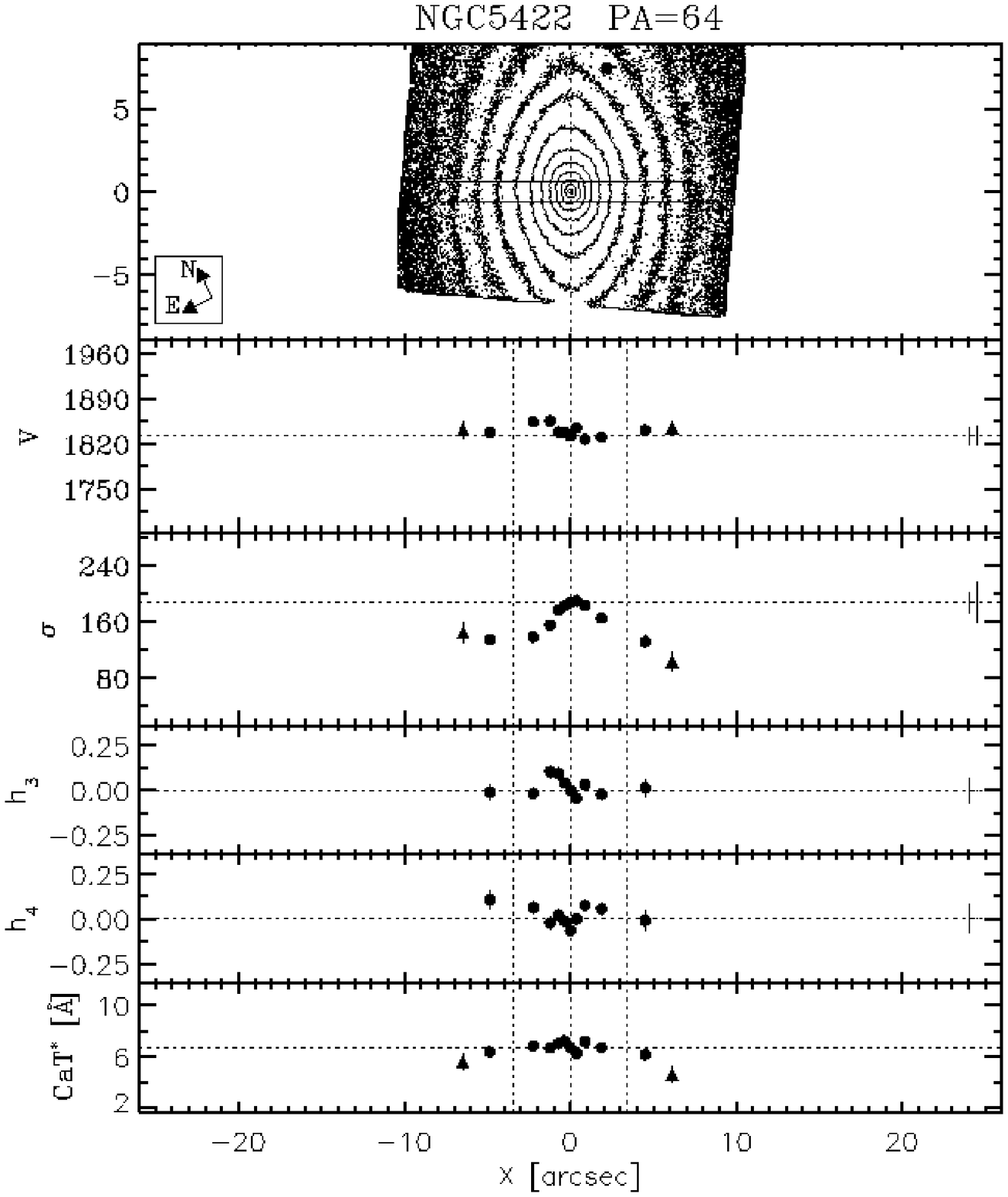}
\hfill%
\includegraphics[height=10.5cm,angle=0]{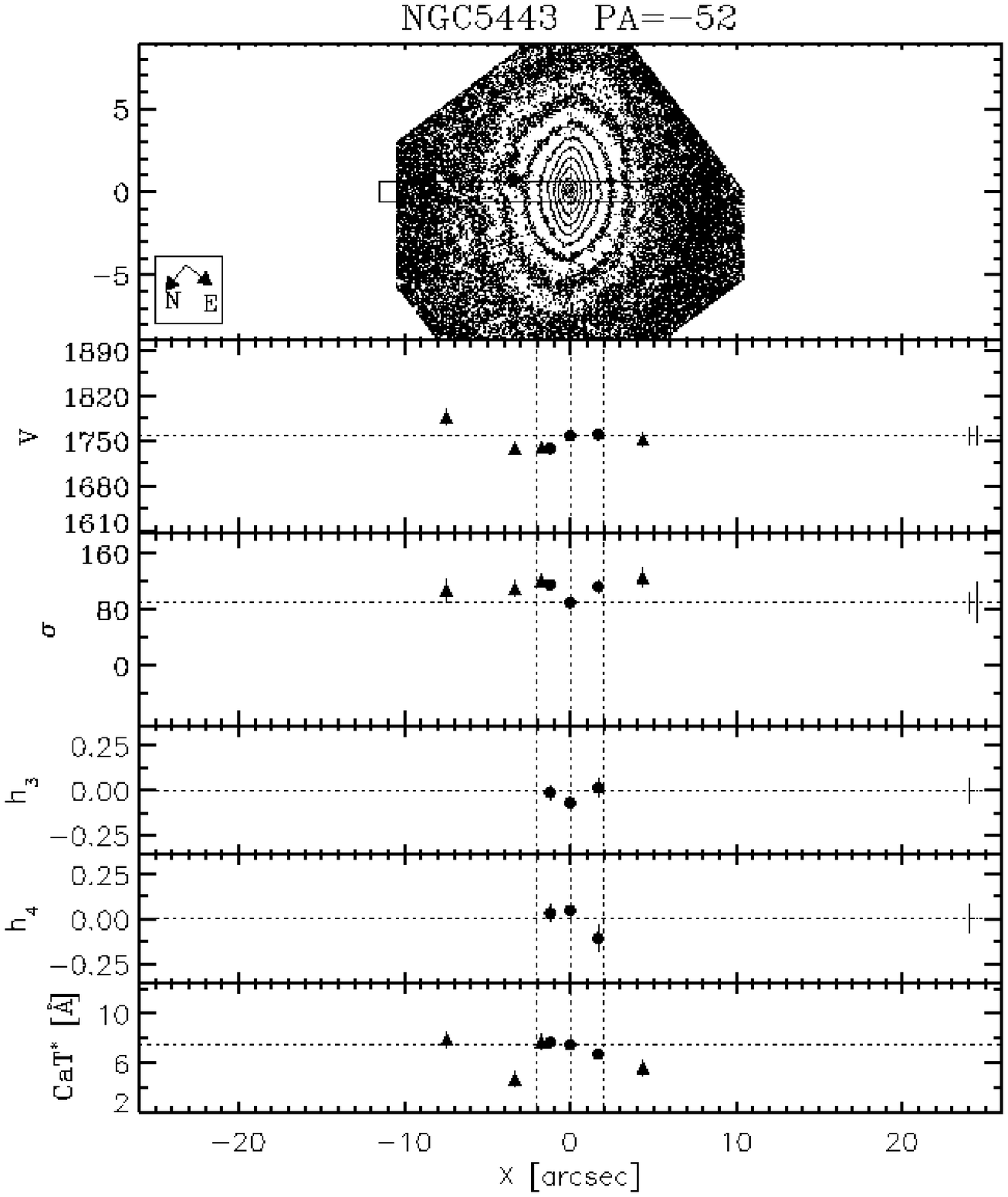}
\end{figure*}

\clearpage

\begin{figure*}
\vspace{2cm}
\centering
\includegraphics[height=10.5cm,angle=0]{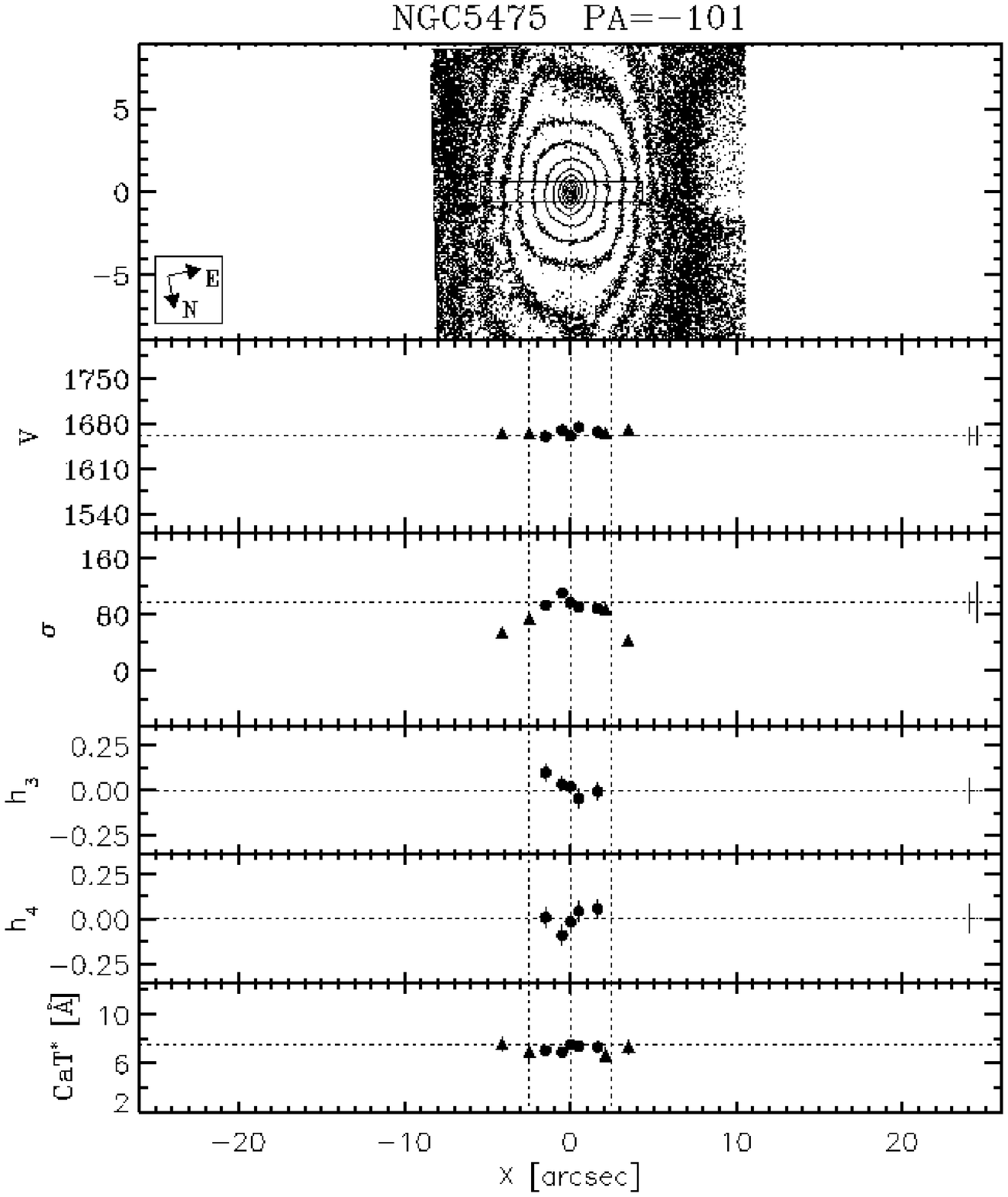}
\hfill%
\includegraphics[height=10.5cm,angle=0]{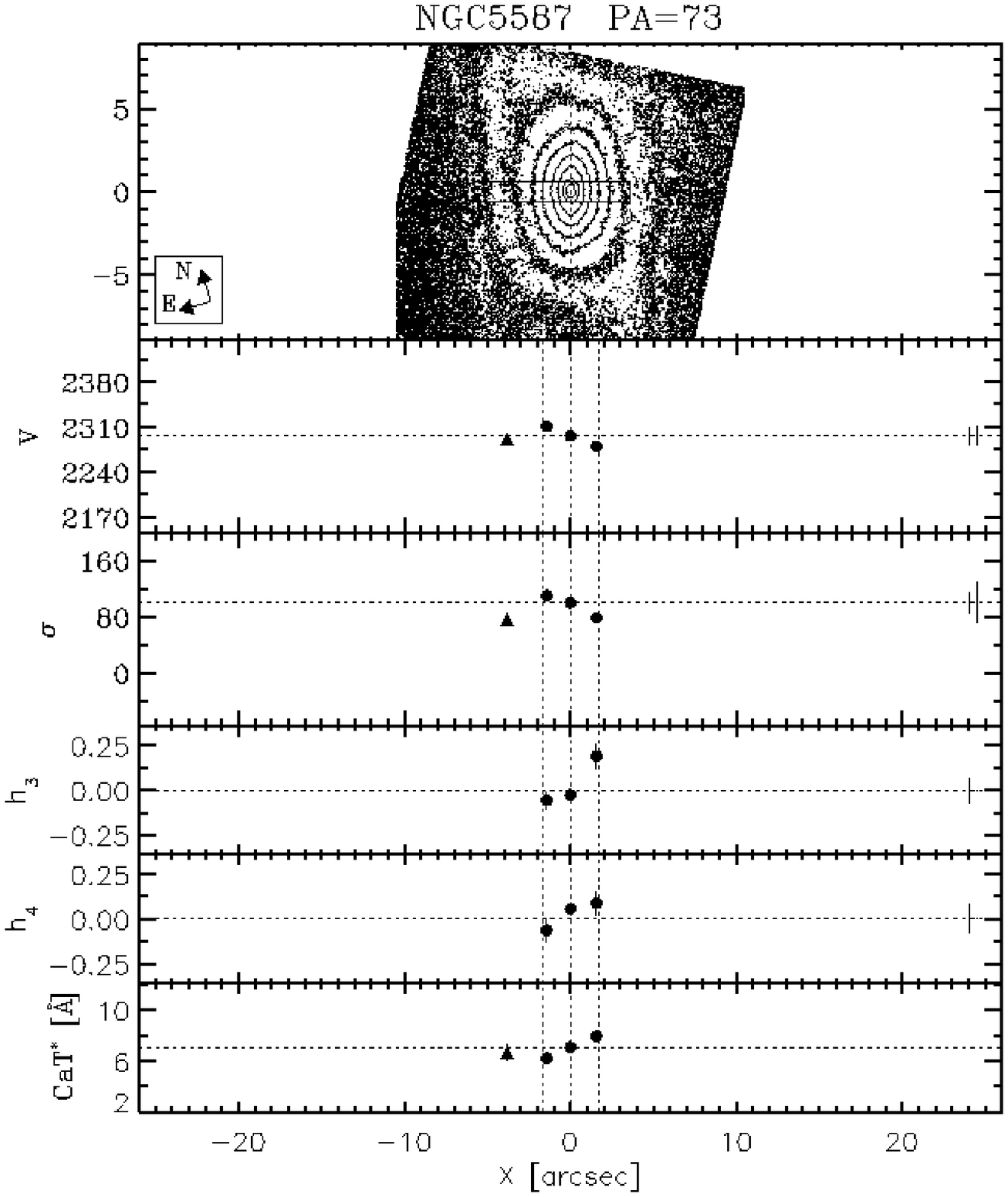}
\vspace{2cm}
\\
\centering
\includegraphics[height=10.5cm,angle=0]{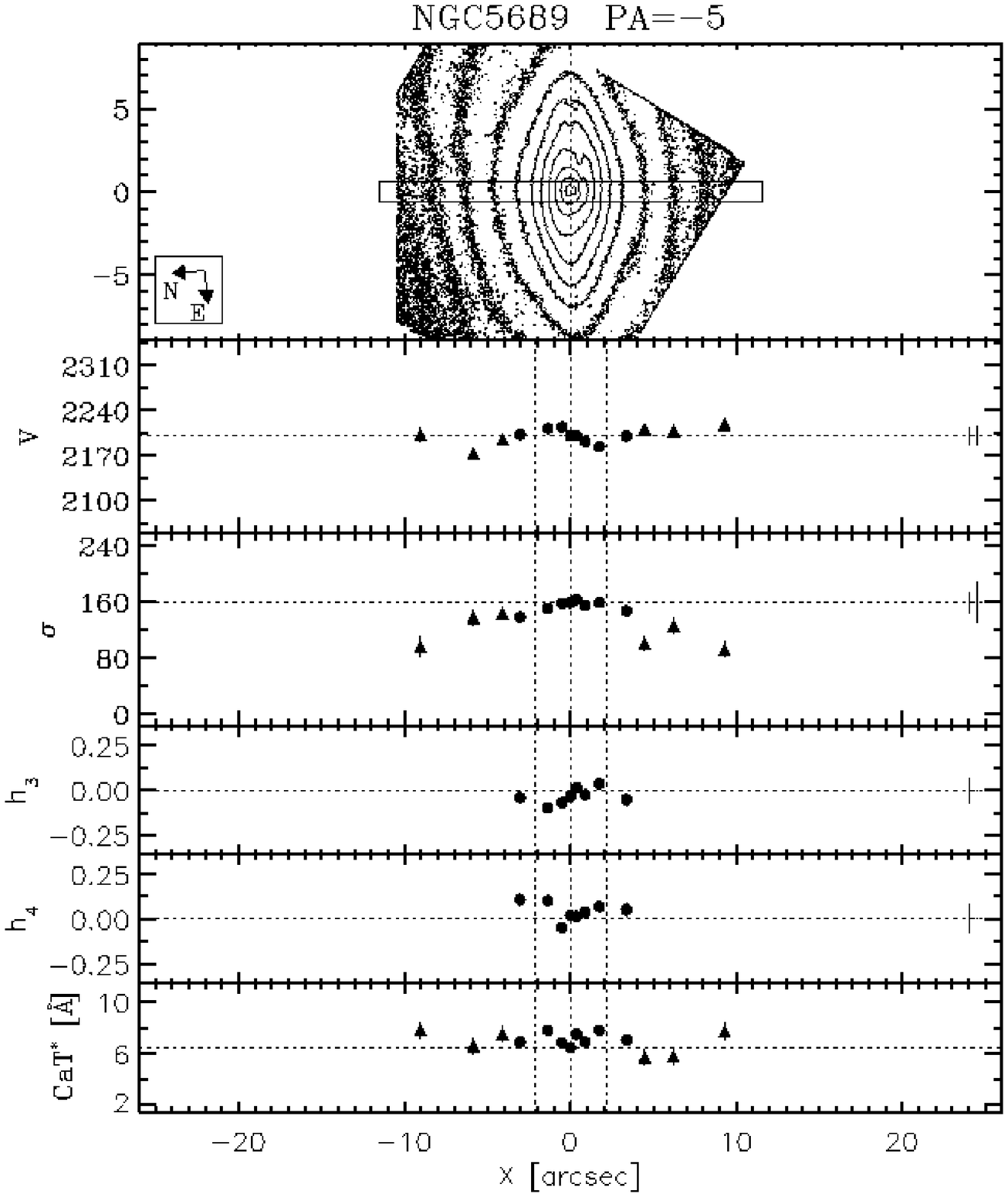}
\hfill%
\includegraphics[height=10.5cm,angle=0]{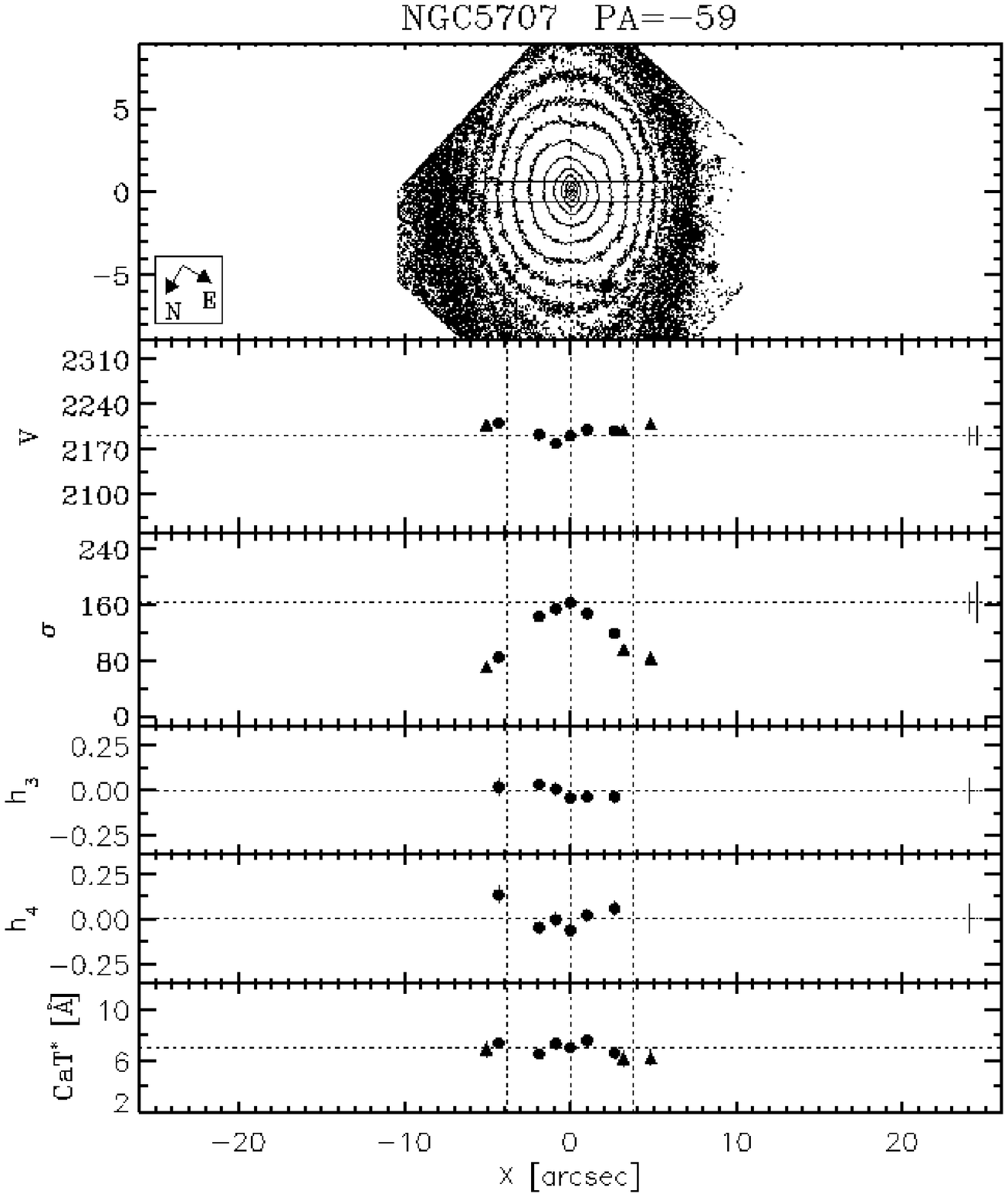}
\end{figure*}

\clearpage

\begin{figure*}
\vspace{2cm}
\centering
\includegraphics[height=10.5cm,angle=0]{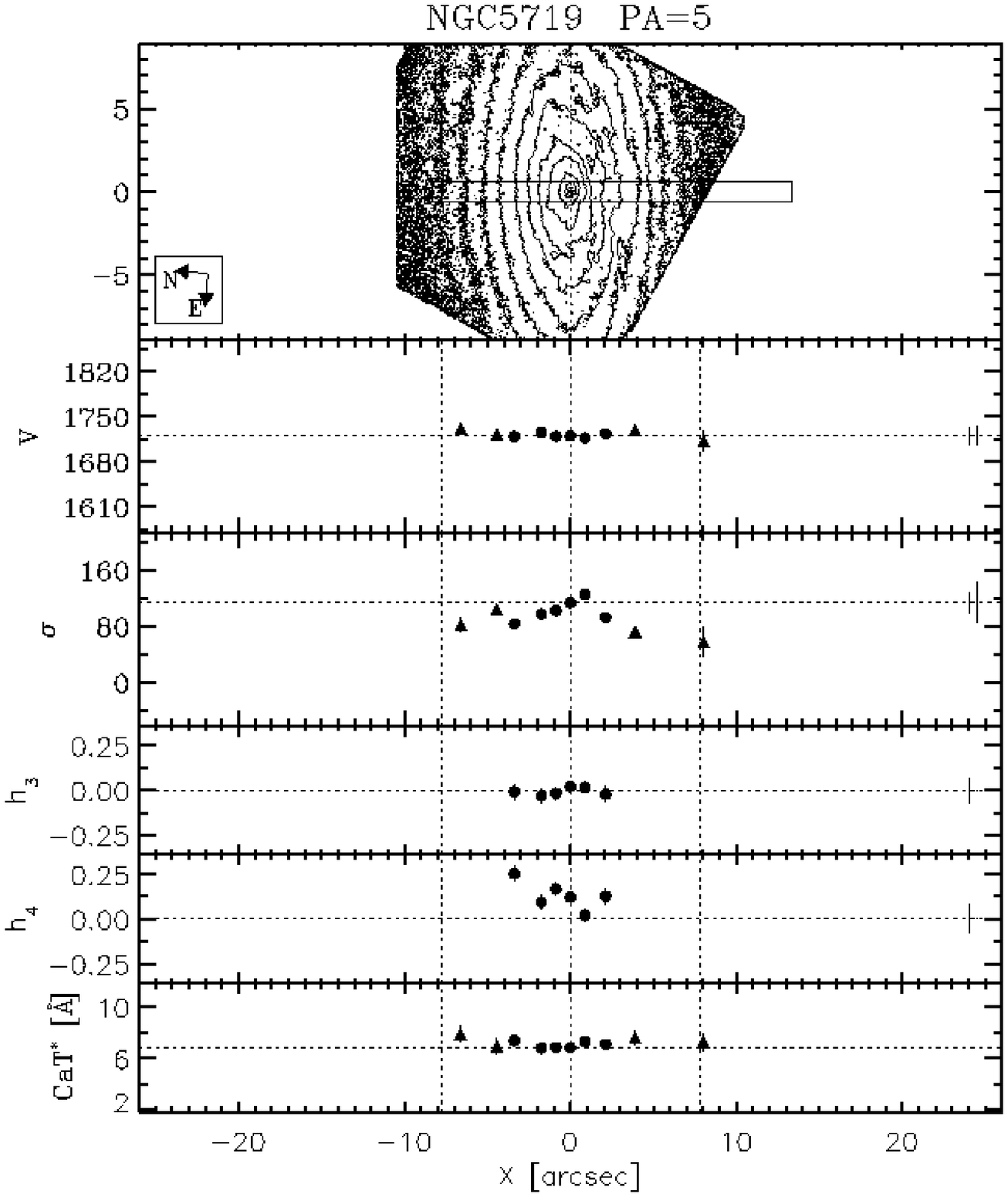}
\hfill%
\includegraphics[height=10.5cm,angle=0]{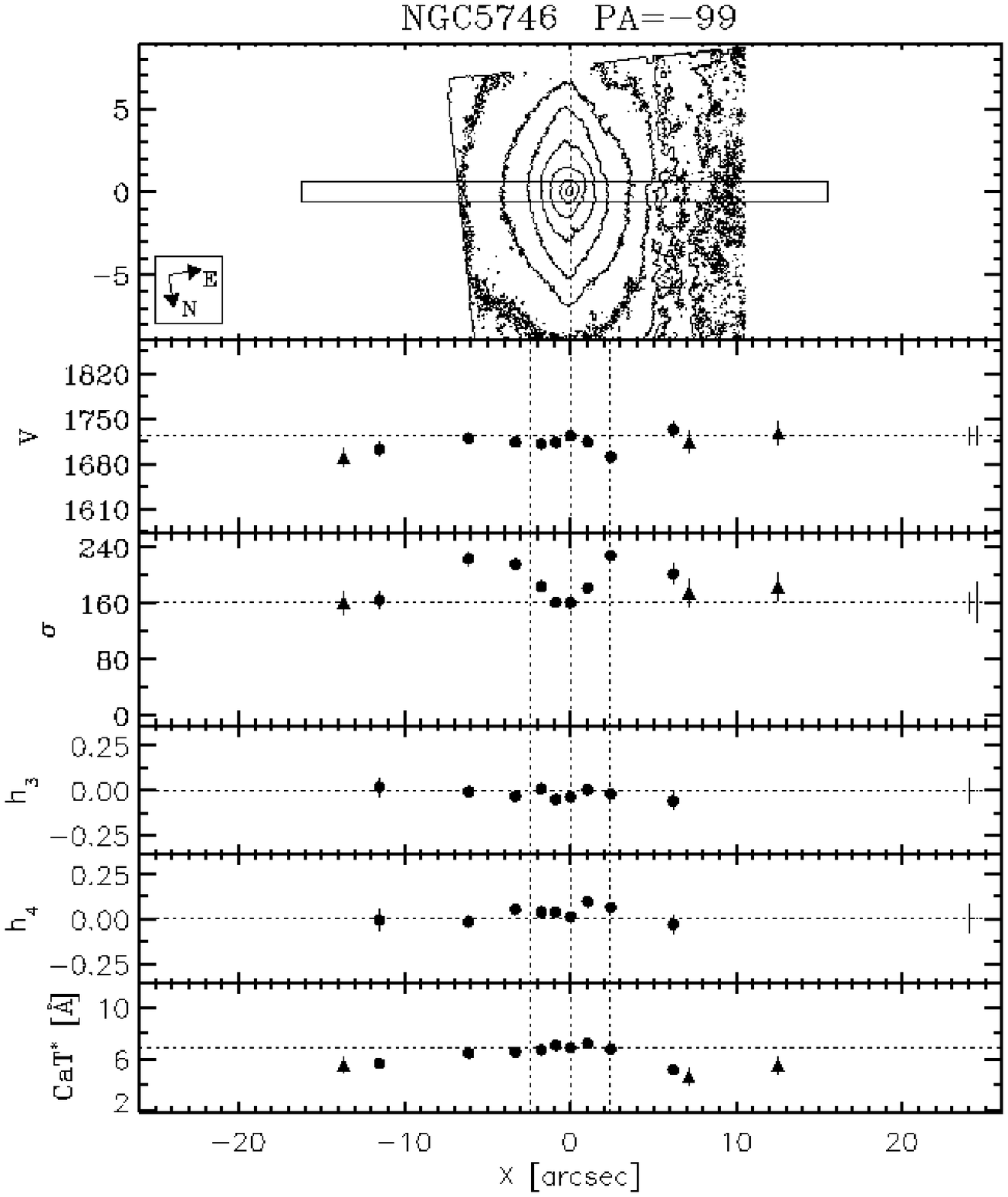}
\vspace{2cm}
\\
\centering
\includegraphics[height=10.5cm,angle=0]{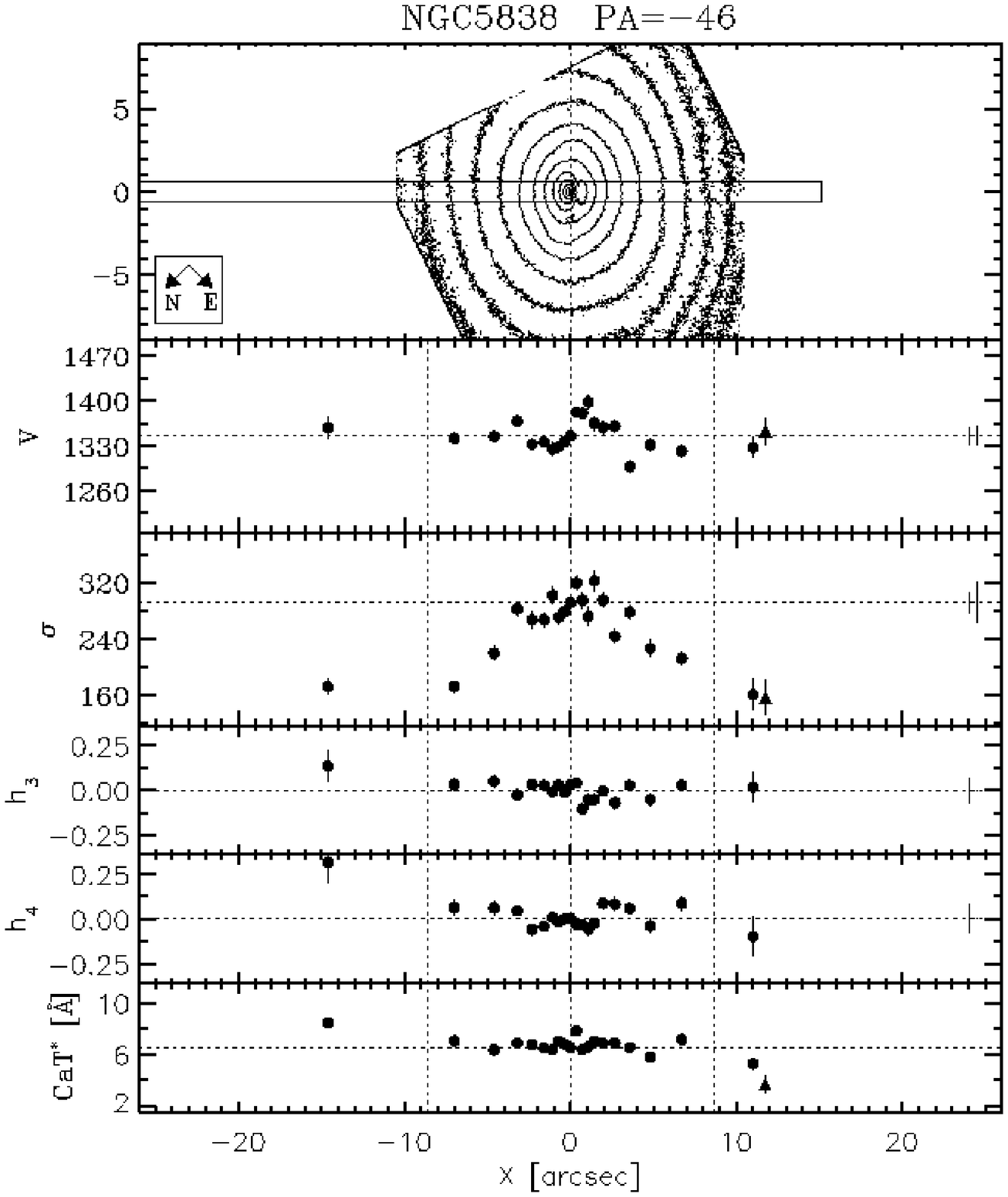}
\hfill%
\includegraphics[height=10.5cm,angle=0]{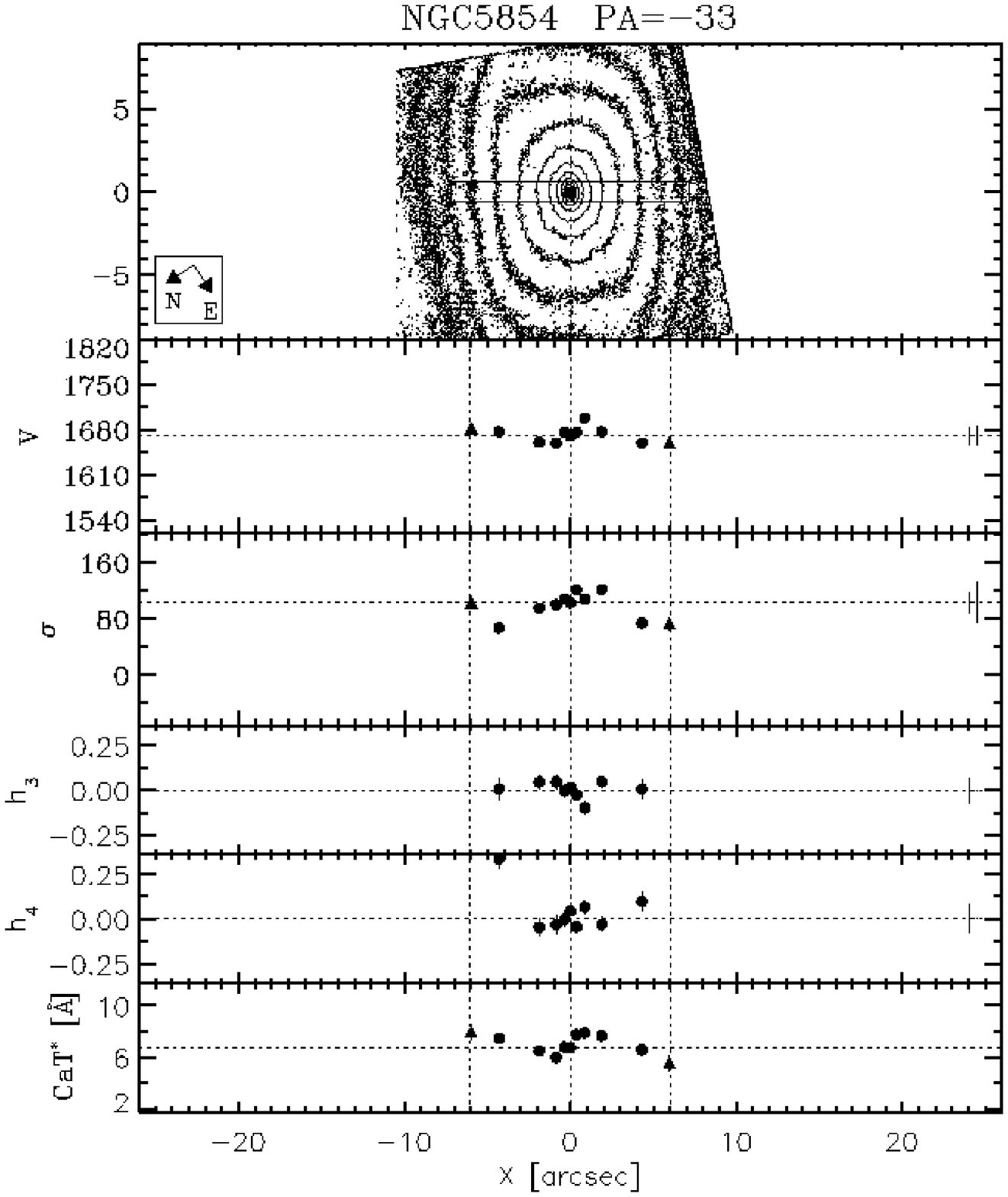}
\end{figure*}

\clearpage

\begin{figure*}
\vspace{2cm}
\centering
\includegraphics[height=10.5cm,angle=0]{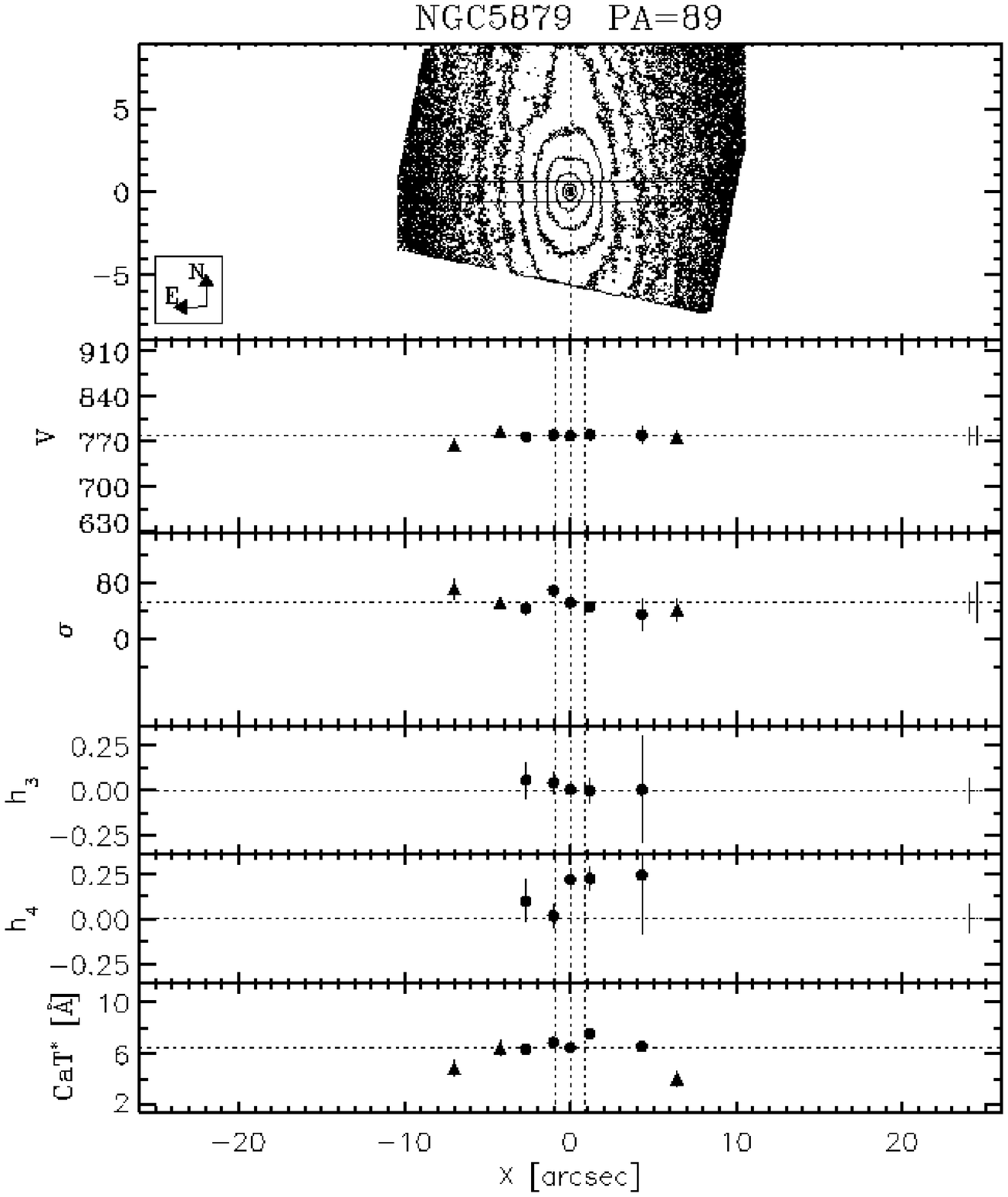}
\hfill%
\includegraphics[height=10.5cm,angle=0]{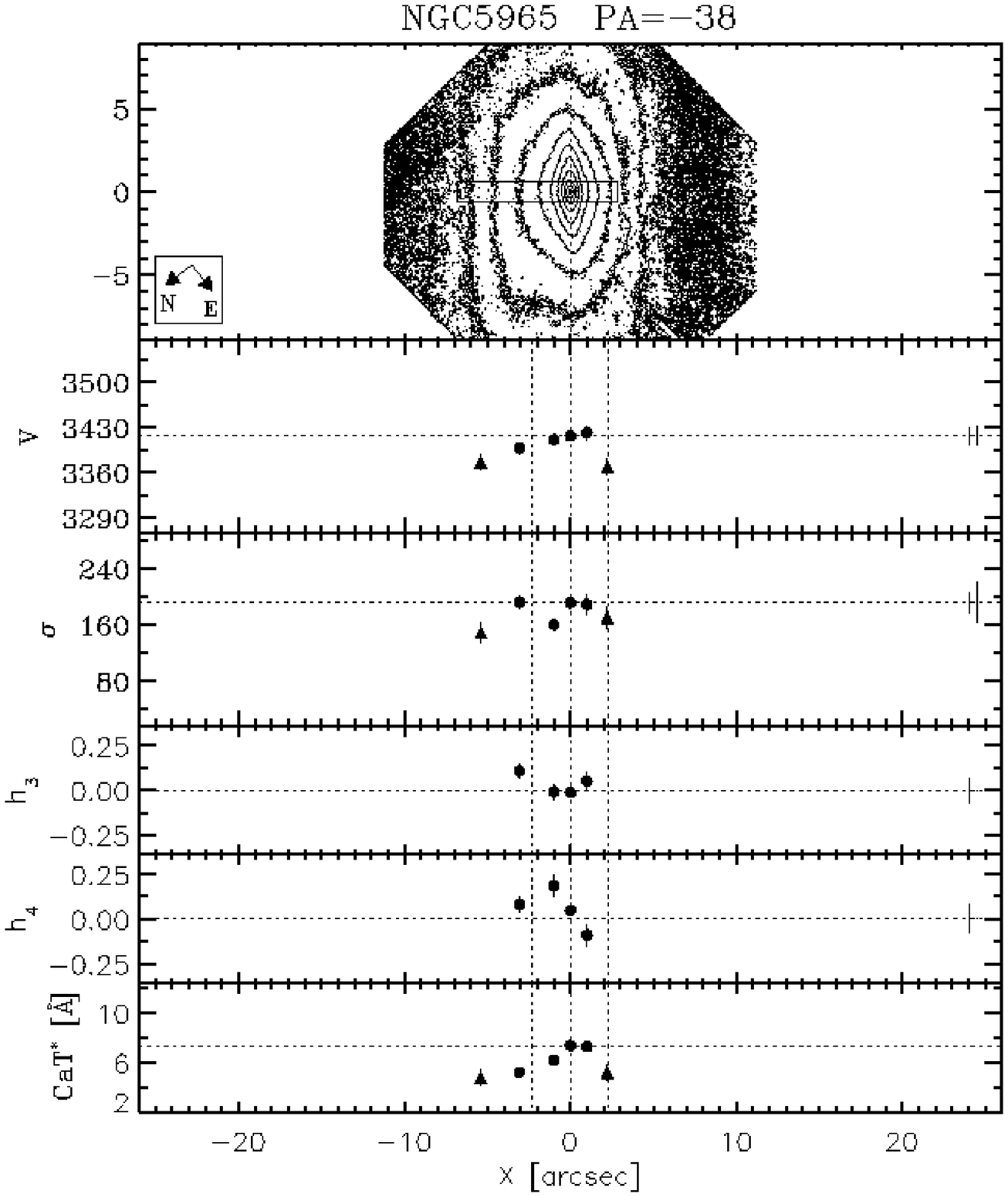}
\vspace{2cm}
\\
\centering
\includegraphics[height=10.5cm,angle=0]{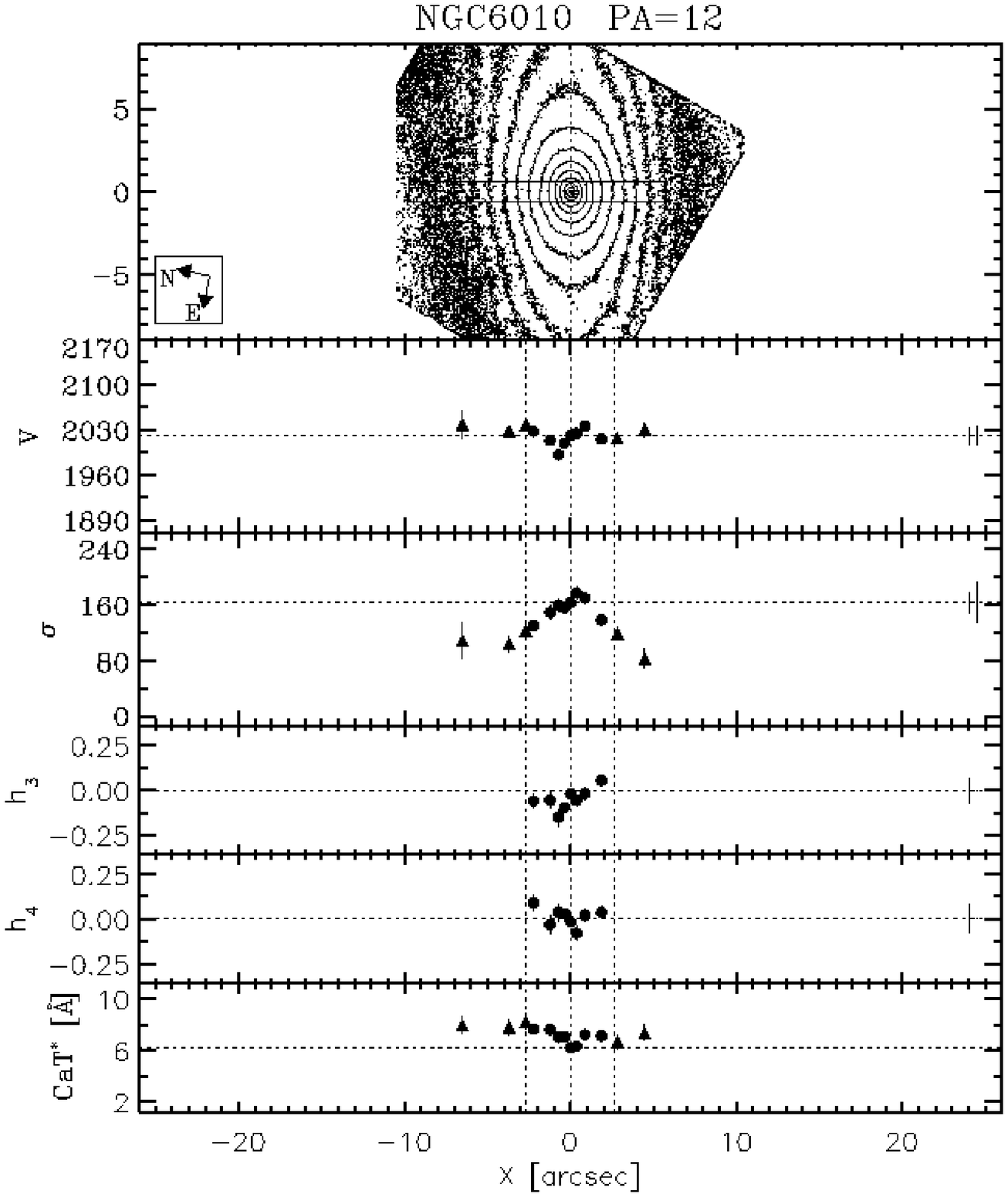}
\hfill%
\includegraphics[height=10.5cm,angle=0]{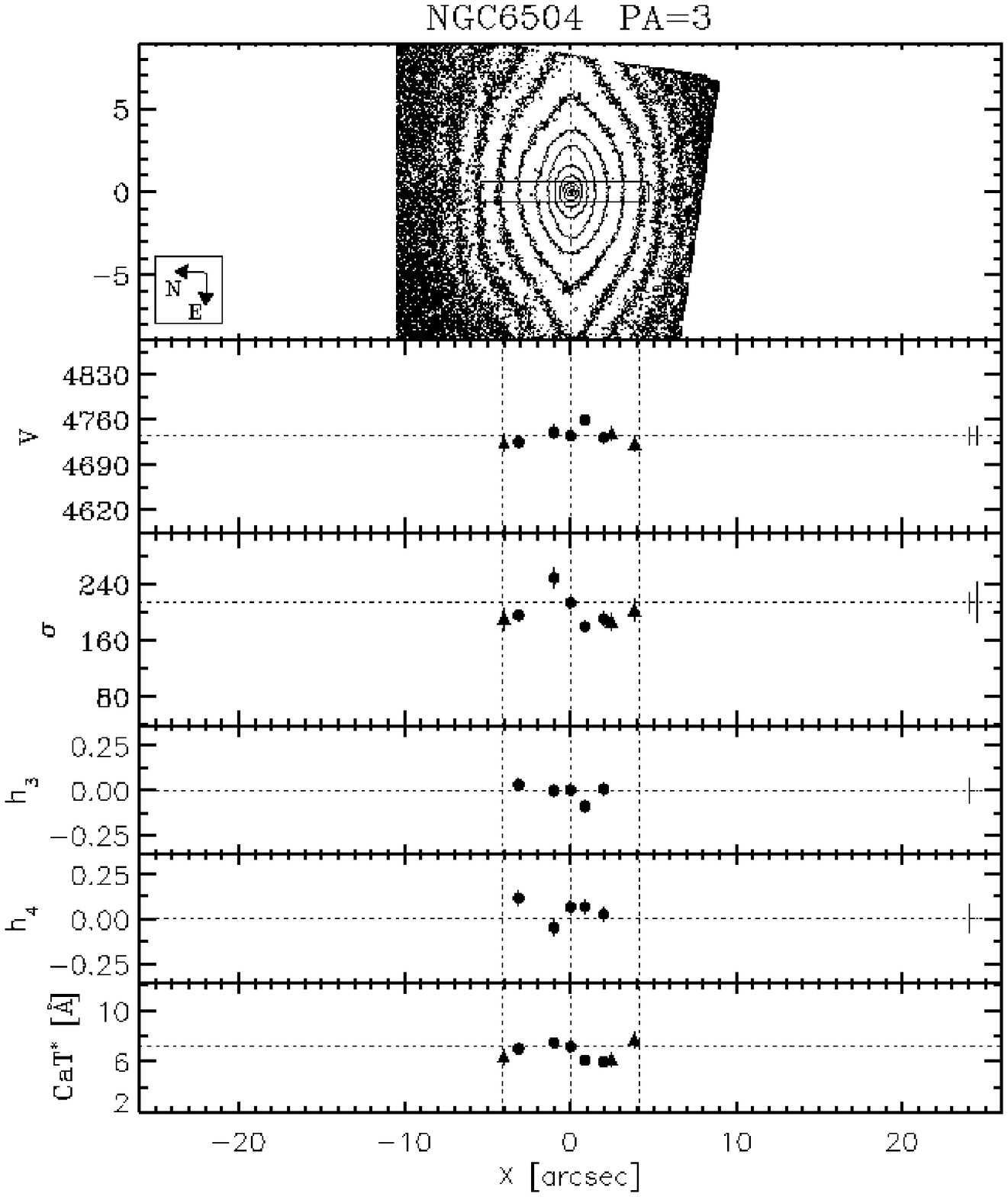}
\end{figure*}

\clearpage

\begin{figure*}
\vspace{2cm}
\centering
\includegraphics[height=10.5cm,angle=0]{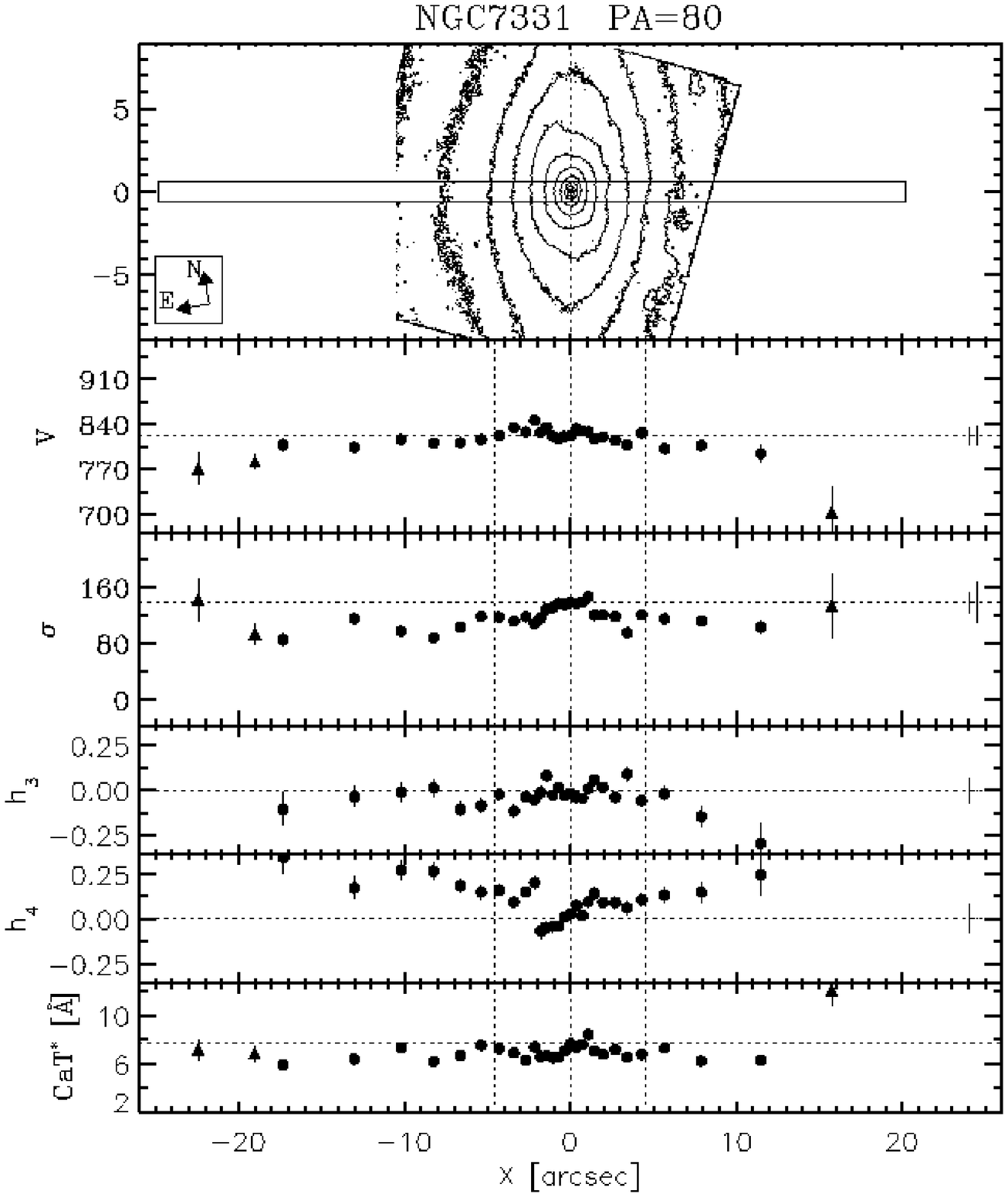}
\hfill%
\includegraphics[height=10.5cm,angle=0]{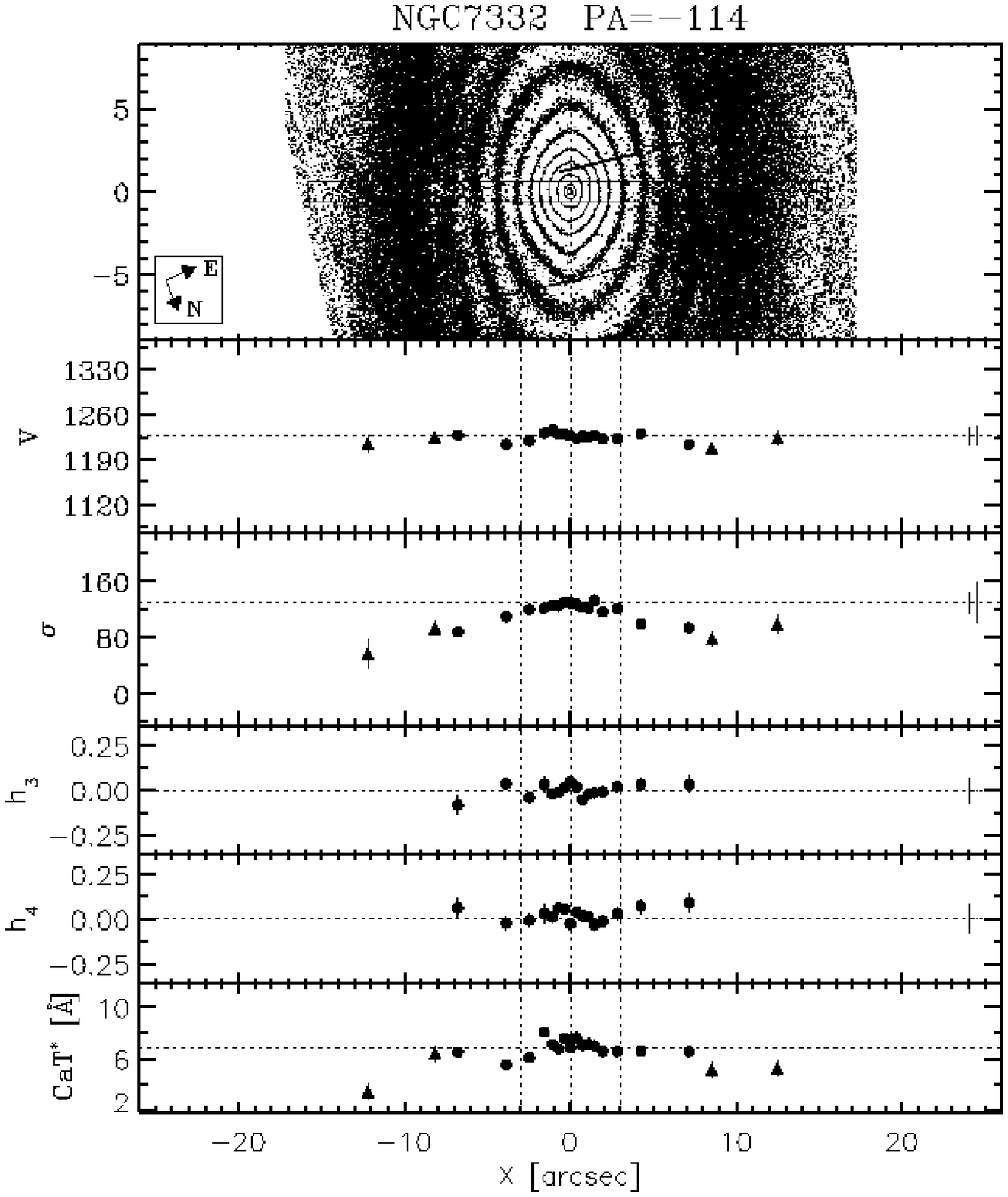}
\vspace{2cm}
\\
\centering
\includegraphics[height=10.5cm,angle=0]{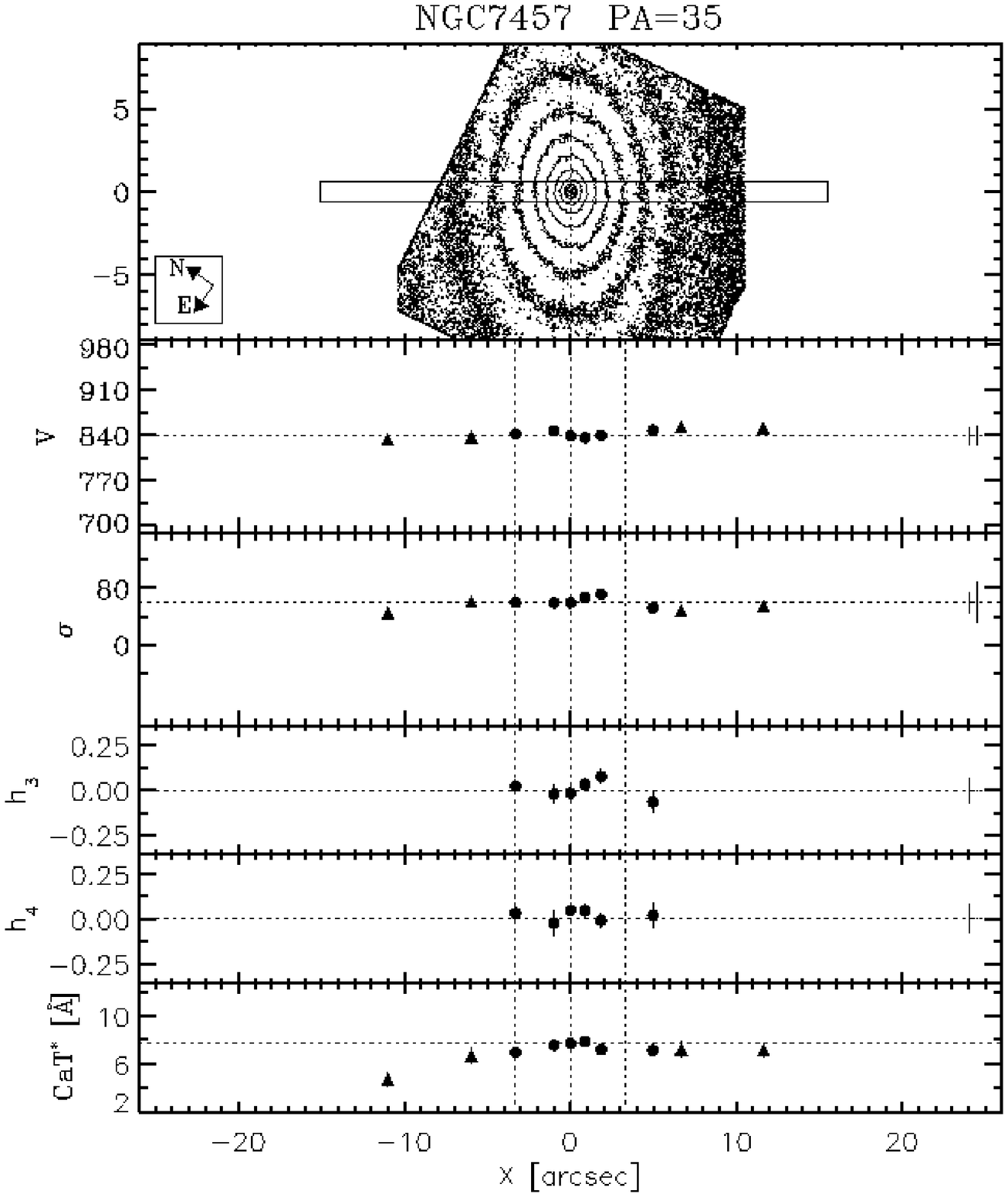}
\end{figure*}

\clearpage

\end{document}